\documentclass[aps,pre,twocolumn,superscriptaddress]{revtex4}
\makeatletter
\@addtoreset{equation}{section}
\makeatother

\usepackage{graphicx}
\usepackage{subfigure}
\makeatletter
\renewcommand{\p@subfigure}{\thefigure}
\makeatother
\usepackage{dcolumn}
\usepackage{bm}
\usepackage[section]{placeins}
\usepackage{morefloats}
\usepackage{color}
\begin{document}

\title{Compressible turbulent mixing: Effects of Schmidt number}
\author{Qionglin Ni}
\email{niql.pku@gmail.com}
\affiliation{State Key Laboratory for Turbulence and Complex Systems, College of Engineering, Peking University, 100871,
Beijing, People's Republic of China}
\affiliation{Department of Physics, University of Rome "Tor Vergata",
Via della Ricerca Scientifica 1, 00133, Rome, Italy}
\date{\today}

\begin{abstract}
We investigated by numerical simulations the effects of Schmidt number on passive scalar transport in forced compressible
turbulence. The range of Schmidt number ($Sc$) was $1/25\sim 25$. In the inertial-convective range the scalar
spectrum was seemed to obey the $k^{-5/3}$ power law. For $Sc\gg 1$, there appeared a $k^{-1}$ power law in the
viscous-convective range, while for $Sc\ll 1$, a $k^{-17/3}$ power law was identified in the inertial-diffusive range.
The scaling constant computed by the mixed third-order structure function of velocity-scalar increment showed
that it grew over $Sc$, and the effect of compressibility made it smaller than the $4/3$ value from incompressible turbulence.
At small amplitudes, the probability distribution function (PDF) of scalar fluctuations collapsed to the Gaussian distribution,
whereas at large amplitudes it decayed more quickly than Gaussian. At large scales, the PDF of scalar increment behaved similarly
to that of scalar fluctuation. In contrast, at small scales it resembled the PDF of scalar gradient. Further, the
scalar dissipation occurring at large magnitudes was found to grow with $Sc$. Due to low molecular
diffusivity, in the $Sc\gg 1$ flow the scalar field rolled up and got mixed sufficiently. However, in the $Sc\ll 1$ flow the scalar
field lost the small-scale structures by high molecular diffusivity, and retained only the large-scale, cloudlike structures. The
spectral analysis found that the spectral densities of scalar advection and dissipation in both $Sc\gg 1$ and $Sc\ll 1$ flows
probably followed the $k^{-5/3}$ scaling. This indicated that in compressible turbulence the processes of advection and
dissipation except that of scalar-dilatation coupling might defer to the Kolmogorov picture. It then
showed that at high wavenumbers, the magnitudes of spectral coherency in both $Sc\gg 1$ and $Sc\ll 1$ flows decayed faster
than the theoretical prediction of $k^{-2/3}$ for incompressible flows. Finally, the comparison with incompressible
results showed that the scalar in compressible turbulence with $Sc=1$ lacked a conspicuous bump structure in its spectrum, but was
more intermittent in the dissipative range.
\end{abstract}
\maketitle

\section{INTRODUCTION}

Turbulent mixing is of importance in many fields including the scattering of interstellar materials throughout the Universe,
the dispersion of air pollutants in the atmosphere, and the combustion of chemical reactions within an engine [1-6]. In the
literature of fluid dynamics, mixing in turbulent flows is often called as scalar turbulence. The related classical
picture of cascade is that scalar fluctuations are generated at large scales and transported through successive
breakdowns into smaller scales, the process proceeds until the scalar fluctuations are homogenized and dissipated by
molecular diffusion at the smallest scale. Therefore, how a scalar gets mixed by a flow depends on whether its molecular
diffusivity is small or large, even if the flow is fully turbulent. The common measure of diffusivity is based on the Schmidt
number $Sc\equiv \nu/\chi$, where $\nu$ and $\chi$ are the kinematic viscosity and molecular diffusivity,
respectively. Generally, there are three different $Sc$ regimes in turbulent mixing.
For $Sc\simeq 1$, previous experiments and simulations [7-9] have suggested that the scalar spectrum in the inertial-convective
range follows the $k^{-5/3}$ power law when the Reynolds number is sufficiently high. The weakly diffusive regime
($Sc\gg 1$) has also received great deal of attention [10-12], especially concerning the $k^{-1}$ power law for the spectral
roll-off in the viscous-convective range [13]. In terms of the strongly diffusive regime ($Sc\ll 1$), recent simulations [14,15]
have provided strong support for the putative theory proposed by Batchelor, Howells and Townsend (hereafter referred to as BHT) [16],
namely, that the scalar spectrum in the inertial-diffusive range obeys the $k^{-17/3}$ power law.

As is well known, compressible turbulence is crucial to a large number of industrial applications and natural phenomena, such as the
design of transonic and hypersonic aircrafts, inter-planet space exploration, solar winds, and star-forming clouds in a galaxy.
Nevertheless, our current understanding of scalar transport in compressible turbulent flows lags far behind the knowledge
accumulated on the incompressible one. Previous simulations of mixing in compressible turbulence using the
piecewise-parabolic method [17,18] showed that for velocity, the compressive component is less efficient in enhancing mixing
than the solenoidal component. Moreover, the scaling of scalar structure function accords well with the SL94 model.
In this paper, we carried out a series of numerical simulations for compressible turbulent mixing, using a novel computational
approach [19]. To examine in detail the effects of the Schmidt number on the scalar transport in compressible turbulence, the
turbulent Mach number was fixed at around $0.30$, whereas the Taylor microscale Reynolds and Schmidt numbers were varied from
$34$ to $216$ and $1/25$ to $25$, respectively. For focusing the influences brought by shock waves, a large-scale forcing with
overwhelming compressive component is added to drive and maintain the velocity field [20]. This paper is part of a systemic
investigation of the effects of basic parameters on compressible turbulent mixing. In a companion paper [21], we carefully
examined the effects caused by changes in the Mach number and forcing scheme.

The rest of this paper is organized as follows. The governing equations and simulated parameter, along with the computational
method used, are presented in Sec. 2. In Sec. 3, we first analyze the spectrum and structure function, then describe
the probability distribution function, and finally discuss the scalar transport in $Sc\gg 1$ and $Sc\ll 1$ flows. The
summary and conclusions regarding this paper are given in Sec. 4.

\section{GOVERNING EQUATIONS AND SIMULATION PARAMETERS}

We consider a statistically stationary system of a passive scalar advected by the compressible turbulence of an idea gas.
The velocity and scalar fields are driven and maintained by large-scale velocity and scalar forcings, respectively, where in
the former the ratio of compressive to solenoidal components for each wavenumber is $20:1$. Furthermore, the accumulated internal
energy at small scales is removed by cooling function at large scales. By introducing the basic scales of $L$ for
length, $\rho_0$ for density, $U$ for velocity, $T_0$ for temperature and $\phi_0$ for scalar, we obtain the dimensionless form
of governing equations, plus the dimensionless state equation of ideal gas, as follows
\begin{equation}
\frac{\partial\rho}{\partial t} + \frac{\partial\big(\rho u_j\big)}{\partial x_j}=0,
\end{equation}
\begin{equation}
\frac{\partial\big(\rho u_i\big)}{\partial t} + \frac{\partial\big[\rho u_iu_j + p\delta_{ij}/\gamma M^2\big]}
{\partial x_j}=\frac{1}{Re}\frac{\partial\sigma _{ij}}{\partial x_j} + \rho{\cal F}_i,
\end{equation}
$$\frac{\partial{\cal E}}{\partial t} + \frac{\partial\big[({\cal E}+ p/\gamma M^2)u_j\big]}{\partial x_j}
=\frac{1}{\alpha}\frac{\partial}{\partial x_j}\big(\kappa\frac{\partial T}{\partial x_j}\big)$$
\begin{equation}
+ \frac{1}{Re}\frac{\partial\big(\sigma _{ij}u_i\big)}{\partial x_j} - \Lambda + \rho{\cal F}_j u_j,
\end{equation}
\begin{equation}
\frac{\partial\big(\rho\phi\big)}{\partial t} + \frac{\partial\big[(\rho\phi)u_j\big]}{\partial x_j}
=\frac{1}{\beta}\frac{\partial}{\partial x_j}\big(\rho\chi\frac{\partial\phi}{\partial x_j}\big) + \rho{\cal S},
\end{equation}
\begin{equation}
p=\rho T.
\end{equation}
The primary variables are the density $\rho$, velocity vector $\textbf{u}$, pressure $p$, temperature $T$ and scalar $\phi$.
The nondimensional parameters $\alpha$ and $\beta$ are $\alpha=PrRe(\gamma-1)M^2$ and $\beta=ScRe(\gamma-1)\gamma$.
${\cal F}_j$ is the dimensionless large-scale velocity forcing
\begin{equation}
{\cal F}_j = \sum_{l=1}^{2}\hat{{\cal F}_j}(\textbf{k}_l)\exp(i\textbf{k}_l\textbf{x})+c.c.,
\end{equation}
where $\hat{{\cal F}_j}$ is the Fourier amplitude, it has a solenoidal component perpendicular to $\textbf{k}_l$ and
a compressive component parallel to $\textbf{k}_l$, and in magnitude the latter is twenty times over the former.
Similarly, the dimensionless large-scale scalar forcing $\cal S$ is written as
\begin{equation}
{\cal S} = \sum_{l=1}^{2}\hat{\cal S}(\textbf{k}_l)\exp(i\textbf{k}_l\textbf{x})+c.c..
\end{equation}
However, there is only a solenoidal component in $\hat{\cal S}$. The details of the thermal cooling function $\Lambda$
can be found in [19]. The viscous stress $\sigma_{ij}$ and total energy per unit volume $\cal E$ are defined by
\begin{equation}
\sigma _{ij} \equiv \mu \big(\frac{\partial u_i}{\partial
x_j}+\frac{\partial u_j}{\partial x_i}\big)-\frac{2}{3}\mu\theta\delta_{ij},
\end{equation}
\begin{equation}
{\cal E} \equiv \frac{p}{(\gamma-1)\gamma M^2}+\frac{1}{2}\rho\big(u_ju_j\big).
\end{equation}
Here, $\theta=\partial u_k/\partial x_k$ is the velocity divergence or dilatation. $M\equiv U/c_0$ is the reference Mach number and
$c_0\equiv\sqrt{\gamma RT_0}$ is the reference sound speed, where $R$ is the specific gas constant, and $\gamma\equiv C_p/C_v$
is the ratio of specific heat at constant pressure $C_p$ to that at constant volume $C_v$. We shall assume that both specific heats are
independent of temperature, which is a reasonable assumption for the air temperature in the simulation of current Mach number [19].
By adding the reference dynamical viscosity $\mu_0$, thermal conductivity $\kappa_0$ and molecular diffusivity $\chi_0$, we obtain three
additional governing parameters: the reference Reynolds number $Re\equiv\rho_0UL/\mu_0$, the reference Prandtl number $Pr\equiv
\mu_0C_p/\kappa_0$, and the reference Schmidt number $Sc\equiv \nu_0/\chi_0$, where $\nu_0\equiv \mu_0/\rho_0$ is the
reference kinematic viscosity. In current study the values of $\gamma$ and $Pr$ are set as $1.4$ and $0.7$, respectively. Thus, there
remain three independent parameters of $M$, $Re$ and $Sc$ to govern the system.
For completion, we employ the Sutherland law to specify the temperature-dependent dynamical viscosity, thermal conductivity
and molecular diffusivity as follows
\begin{equation}
\mu, \kappa, \chi = \frac{1.4042T^{1.5}}{T+0.4042}.
\end{equation}

\begin{table*}[t!]
\caption{Flow statistics in the simulations.}
\begin{center}
\small
\begin{tabular*}{0.95\textwidth}{@{\extracolsep{\fill}}rccccccccccccc}
\hline\hline
Case &$Sc$ &$M_t$ &$Re_{\lambda}$ &$\eta$ &$\eta_B$ &$\eta_{OC}$ &$u'$ &$\phi'$ &$E_K$ &$E_\phi$ &$\langle\epsilon\rangle$ &$\langle\epsilon_\phi\rangle$ &$r_\phi$\\
\hline
C1 &$25$ &$0.28$ &$35$ &$0.027$ &$0.005$ &$\times$ &$2.06$ &$2.75$ &$2.13$ &$3.81$ &$1.20$ &$0.80$ &$0.37$  \\
C2 &$5$ &$0.28$ &$35$ &$0.027$ &$0.012$ &$\times$ &$2.06$ &$2.44$ &$2.15$ &$3.00$ &$1.19$ &$0.87$  &$0.52$  \\
C3 &$1$ &$0.28$ &$34$ &$0.027$ &$0.027$ &$\times$ &$2.06$ &$2.14$ &$2.11$ &$2.30$ &$1.21$ &$0.92$  &$0.70$  \\
C4 &$1$ &$0.29$ &$208$ &$0.007$ &$\times$ &$0.007$ &$2.11$ &$2.24$ &$2.28$ &$2.53$ &$0.41$ &$0.90$  &$1.98$  \\
C5 &$1/5$ &$0.29$ &$207$ &$0.007$ &$\times$ &$0.023$ &$2.11$ &$2.11$ &$2.24$ &$2.25$ &$0.40$ &$0.93$  &$2.31$  \\
C6 &$1/25$ &$0.30$ &$216$ &$0.007$ &$\times$ &$0.078$ &$2.15$ &$1.98$ &$2.30$ &$1.97$ &$0.43$ &$1.25$  &$3.40$  \\
\hline\hline
\end{tabular*}
\normalsize \label{tableI}
\end{center}
\end{table*}

The system is solved numerically in a cubic box with periodic boundary conditions, by adopting a new computational
method. This method utilizes a seventh-order weighted essentially non-oscillatory (WENO) scheme [22] for shock regions
and an eighth-order compact central finite difference (CCFD) scheme [23] for smooth regions outside shocks. A flux-based
conservative formulation is implemented to optimize the treatment of interface between the two regions and then improve
the computational efficiency. The details have been described in [19]. Instead of $M$ and $Re$, the compressible flow
is directly governed by the turbulent Mach number ($M_t$) and Taylor microscale Reynolds number ($Re_\lambda$) [24],
which are defined as follows
\begin{equation}
M_t\equiv M\frac{u'}{\langle{\sqrt{T}}\rangle},
\end{equation}
\begin{equation}
Re_{\lambda}\equiv Re\frac{u'\lambda\langle\rho\rangle}{\sqrt{3}\langle\mu\rangle},
\end{equation}
where $u'\equiv\sqrt{\left\langle u_j^2\right\rangle}$ and $\lambda\equiv u'/\sqrt{\langle(\partial u_j/\partial x_j)^2\rangle}$
are the root-mean-square (r.m.s.) velocity magnitude and the Taylor microscale, respectively. The sign $\langle\cdot\rangle$
denotes ensemble average and the repetition on subscript stands for Einstein summation.
Here, we point out that the definitions of $M_t$ and $Re_\lambda$ are based on nondimensional variables, and this way is
continuously used in the following text if there is no special illustration.

Table~\ref{tableI} presents the major simulation parameters. The simulations are conducted on a $N^3=512^3$ grid and divided into
two groups according to the value of $Re_\lambda$. The first group including C1, C2 and C3 is used to study scalars with low
molecular diffusivity in low $Re$ flows, where $Sc$ is decreased from $25$, $5$ to $1$, and $Re_\lambda$ is around $35$.
For the three cases, the smallest scale for velocity is the Kolmogorov scale
$\eta\equiv [\langle\mu/(Re\rho)\rangle^3/<\epsilon/\rho>]^{1/4}$, and that for scalar is the Batchelor scale
$\eta_B=Sc^{-1/2}\eta$ with values of $0.005$, $0.012$ and $0.027$. In the second group addressing C4, C5 and C6, we pay our
attention to scalars with high molecular diffusivity in high $Re$ flows, where $Sc$ is decreased
from $1$, $1/5$ and $1/25$, and $Re_\lambda$ is around $210$. Here, the smallest scale for a scalar is
the Corrsin scale, $\eta_C=Sc^{-3/4}\eta$, instead of the Batchelor scale, $\eta_B$, with values of $0.007$, $0.023$ and $0.078$.

The r.m.s. magnitude of velocity $u'$ and the kinetic energy per unit volume $E_K$ are increased by $Re_\lambda$
rather than by $Sc$. In contrast, the r.m.s. magnitude of scalar $\phi'$ and the scalar variance per unit volume
$E_\phi$ mainly grow with $Sc$ in both high and low $Re$ flows, where the related definitions are
$\phi' \equiv \sqrt{\langle\phi^2\rangle}$, $E_K \equiv \langle\rho u_j^2\rangle/2$ and $E_\phi \equiv \langle\rho\phi^2\rangle/2$.
Because $M_t$ is fixed, the ensemble-average value of the kinetic energy dissipation rate $\langle\epsilon\rangle$
becomes only dependent of $Re_\lambda$. In contrast,
that of the scalar dissipation rate $\langle\epsilon_\phi\rangle$ in both high and low $Re$ flows increases as
$Sc$ decreases. Here, $\epsilon\equiv\sigma_{ij}S_{ij}/Re$, $\epsilon_\phi\equiv\chi(\partial\phi/\partial x_j)^2$,
and $S_{ij}=({\partial u_i/\partial x_j+\partial u_j/\partial x_i})/2$ is the stain rate tensor.
For the ratio of the mechanical to scalar timescales $r_\phi=(\langle\rho\textbf{u}^2/2\rangle/
\langle\epsilon\rangle)/(\langle\rho\phi^2/2\rangle/\langle\epsilon_\phi\rangle)$, its dependence on $Sc$ is similar
to that of $\langle\epsilon_\phi\rangle$.

\begin{figure}
\centerline{\includegraphics[width=6.5cm]{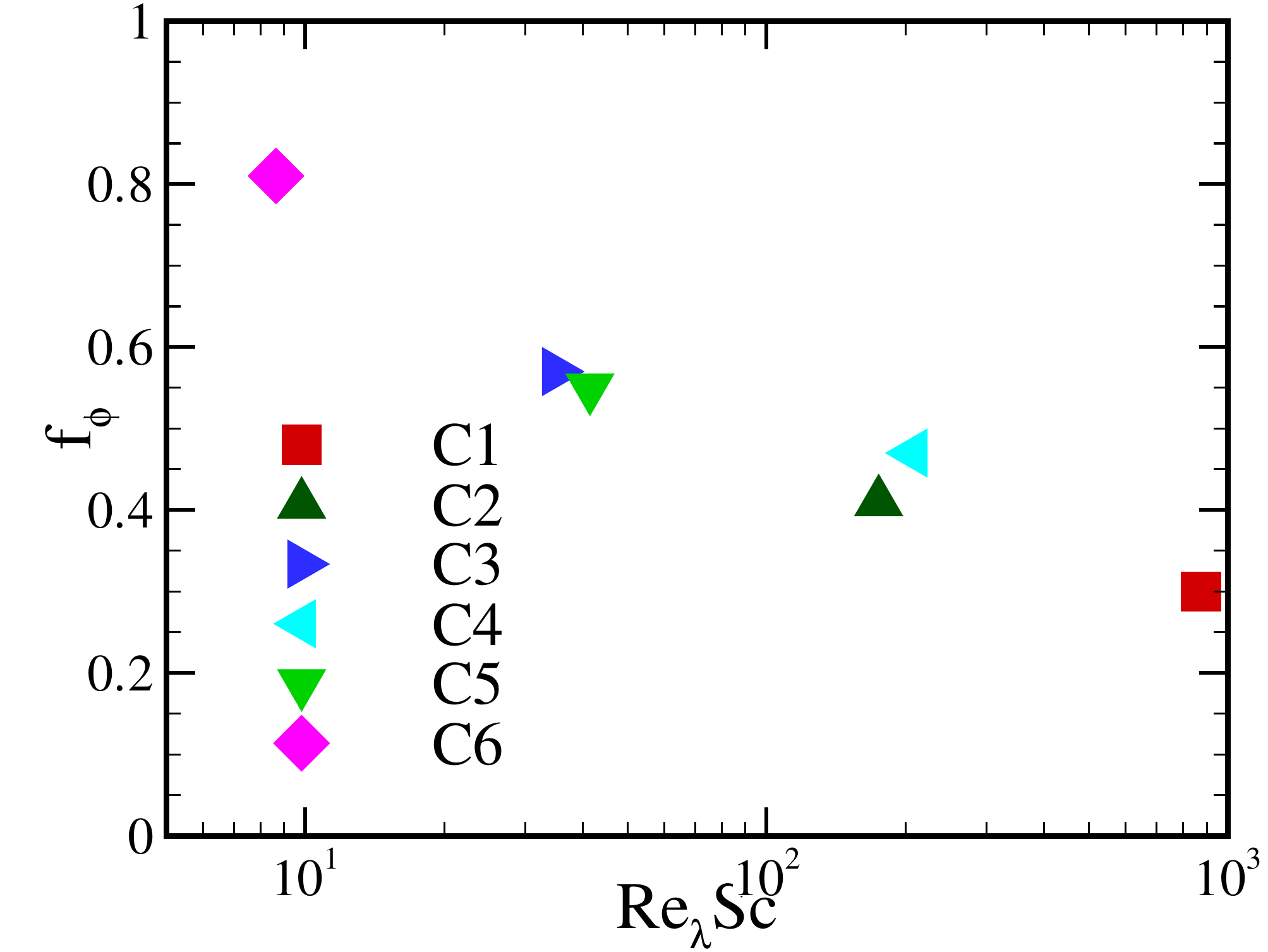}}
\caption{(color online). Normalized scalar dissipation rate versus the product of $Re_\lambda$ and $Sc$}.
\label{fig:fig1}
\end{figure}

Further, an alternative to $r_\phi$ is the quantity of $f_\phi=(L_f/u')/(\langle\rho\phi^2/2\rangle/\langle\epsilon_\phi\rangle)$,
where $L_f$ is the integral length scale [25]. In Fig.~\ref{fig:fig1} we plot $f_\phi$ against the logarithm of the
product of $Re_\lambda$ and $Sc$. It shows that $f_\phi$ falls monotonously when the product increases.

\section{SIMULATION RESULTS}

\subsection{\emph{Spectrum and Structure Function}}

\begin{figure}
\centerline{\includegraphics[width=6.5cm]{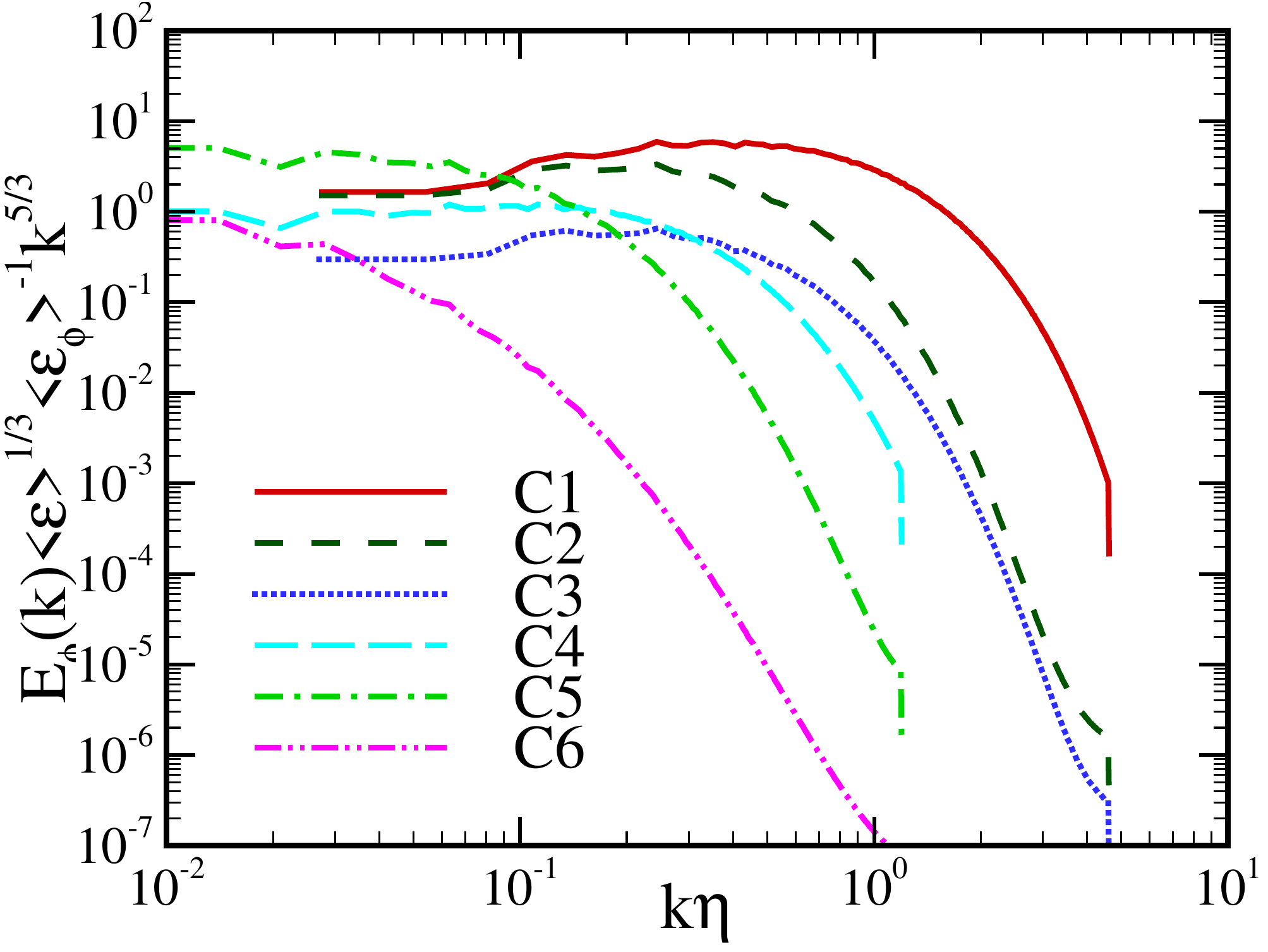}}
\caption{(color online). Compensated spectrum of scalar according to the Obukhov-Corrsin variables at different values of $Sc$ and $Re_\lambda$.}
\label{fig:fig2}
\end{figure}

\begin{figure}
\centerline{\includegraphics[width=6.5cm]{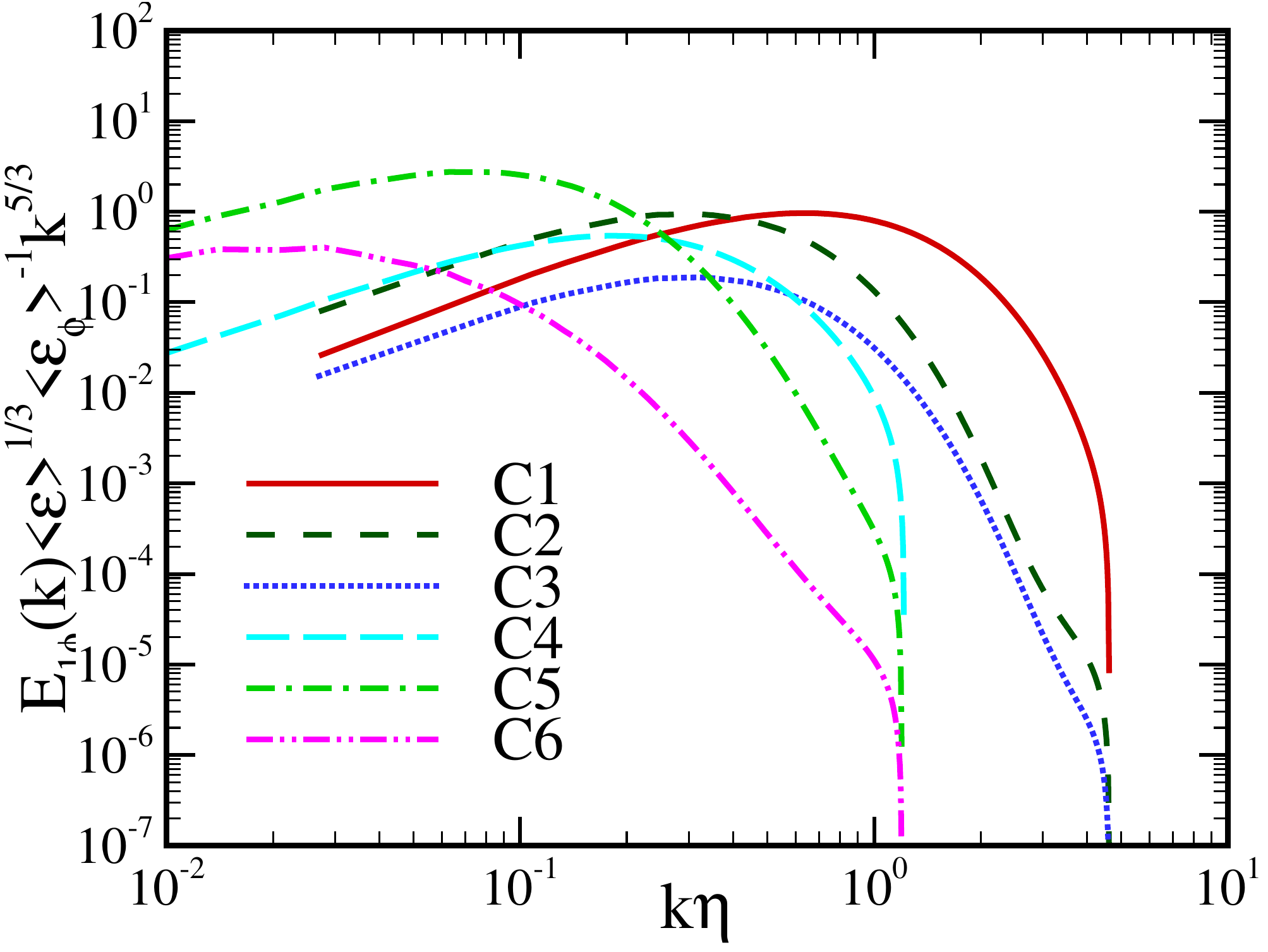}}
\caption{(color online). One-dimensional compensated spectrum of scalar at different values of $Sc$ and $Re_\lambda$.}
\label{fig:fig3}
\end{figure}

By applying the Kolmogorov theory [26,27] to the scalar transport in incompressible turbulent flows, Obukhov [28] and Corrsin
[29] derived a scalar spectrum in the inertial-convective range satisfying
$L_\phi^{-1}\ll k\ll \eta^{-1}$
\begin{equation}
E_{\phi}(k) = C_\phi\langle\epsilon_\phi\rangle \langle\epsilon\rangle^{-1/3} k^{-5/3},
\end{equation}
where $L_\phi$ is the integral length scale of scalar [25]. $C_\phi$ is the Obukhov-Corrsin (OC) constant, and the typical values
are $0.75 \sim 0.92$ by experiments and $0.87\pm 0.10$ by simulations [8,30]. In Fig.~\ref{fig:fig2} we plot the compensated
spectra of scalar according to the OC variables at different $Sc$ and $Re_\lambda$. For the curves of $E_\phi(k)$,
plateaus appear in the inertial-convective ranges, especially for the $Sc=1$ flows. This means that in the range of
$L_\phi^{-1}\ll k\ll \eta^{-1}$, the scalar spectrum in compressible turbulent mixing seems to also obey the $k^{-5/3}$ power law.
In high $Sc$ flows, $E_\phi(k)$ in the regime between the inertial-convective and dissipative ranges
exhibit as growing with wavenumber, which is reinforced by the increase in $Sc$. Contrarily,
in low $Sc$ flows, in the same regime $E_\phi(k)$ falls as wavenumber increases, and this behavior is enhanced by the decrease in $Sc$.
It implies that in a certain range, the scalar spectrum in a low or high $Sc$ flow may have additional scaling.

Although it is straightforward to compute in simulations the three-dimensional (3D) scalar spectrum as a function of wavenumber,
experiments usually measure only the one-dimensional (1D) version of $E_{1\phi}(k)$. In isotropic turbulence, it is written as
\begin{equation}
E_{1\phi}(k) = -\int\limits_k^{\infty}\frac{E_\phi(k)}{k}dk.
\end{equation}
Fig.~\ref{fig:fig3} presents the 1D compensated scalar spectra at different $Sc$ and $Re_\lambda$. We have
taken averages over three coordinate directions. Previous studies of incompressible turbulence have showed the existence
of a spectral bump, which is a precursor to the $k^{-1}$ part of scalar spectrum, and becomes more and more pronounced as $Sc$
increases. In our simulations, although it gets clearer when $Sc$ grows, the bump is not as conspicuous as that observed
in [31]. Similar to $C_{\phi}$, the 1D OC constant, $C_{1\phi}$, is changed by both $Sc$ and $Re_\lambda$.
We find that $C_\phi$ and $C_{1\phi}$ approximately satisfy the relation $C_{\phi}=5C_{1\phi}/3$, which can be obtained directly
through Eq. (3.2).

\begin{figure}
\begin{center}
\subfigure{
\resizebox*{6.5cm}{!}{\rotatebox{0}{\includegraphics{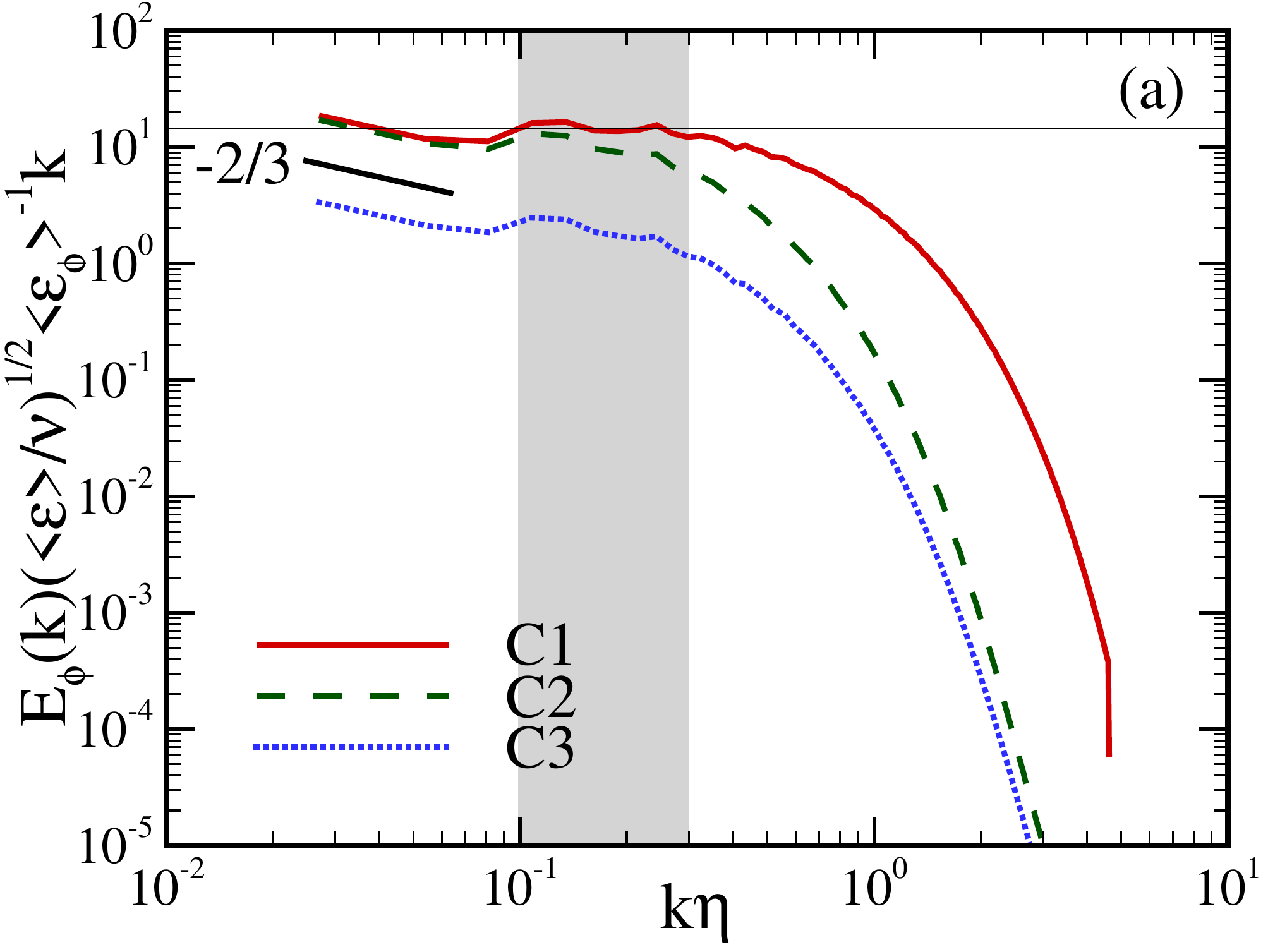}}}}%

\subfigure{
\resizebox*{6.5cm}{!}{\rotatebox{0}{\includegraphics{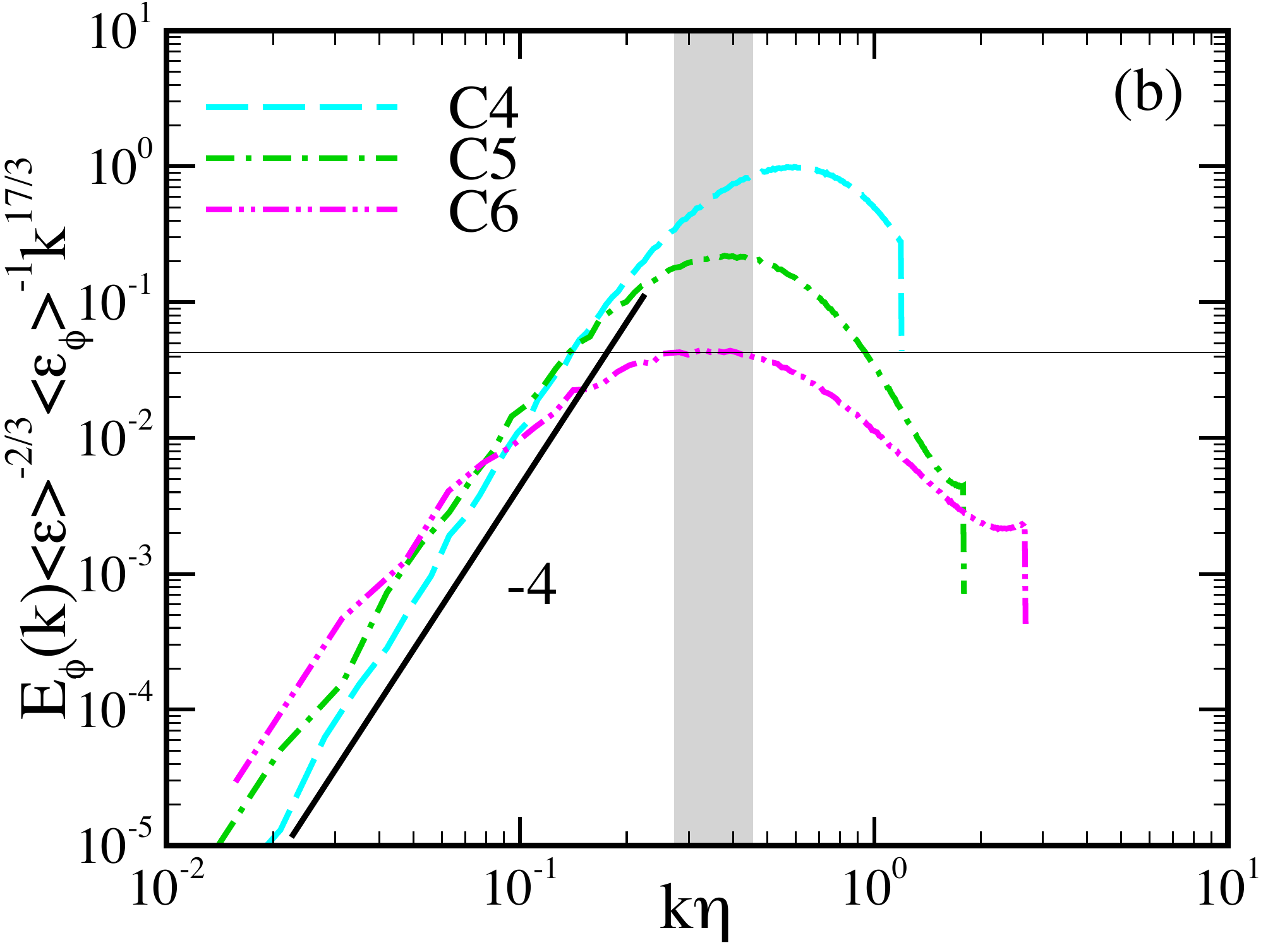}}}}%
\caption{(color online). (a) Compensated spectrum of scalar according to the Batchelor variables at $Sc=25$, $5$ and $1$ and
low $Re_{\lambda}$, where the slope value of the short line is $-2/3$. (b) The same as (a) according to the Batchelor-Howells-Townsend
variables at $Sc=1$, $1/5$ and $1/25$ and high $Re_{\lambda}$, where the slope value of line is $-4$.}
\label{fig:fig4}
\end{center}
\end{figure}

When the Schmidt number is $Sc\gg 1$, the energy spectrum decays quickly at wavenumbers larger than $\eta^{-1}$, whereas the
scalar spectrum remains excited at levels higher than the energy spectrum [32]. In this case, the scalar transfer to small scales
is creased at the Batchelor scale, $\eta_B$, through molecular diffusion. The range $\eta^{-1}\ll k\ll\eta_B^{-1}$ is called as
the viscous-convective range, wherein the scalar spectrum obeys a $k^{-1}$ power law as follows
\begin{equation}
E_{\phi}(k) = B_\phi\langle\epsilon_\phi\rangle\big(\nu/\langle\epsilon\rangle\big)^{1/2}k^{-1}.
\end{equation}
Here, the nondimensional coefficient $B_\phi$ is presumed to be universal [31,32], and the value is $3\sim 6$ [33,34].
The nonlocality of the scalar transfer in wavenumber space is essential for the generation of this viscous-convective
range. Therefore, a sufficiently high $Sc$ is required to observe the $k^{-1}$ power law. In Fig.~\ref{fig:fig4}(a) we
plot the compensated spectra of scalar according to the Batchelor variables in low $Re$ flows, as functions
of $k\eta$. For the $Sc=25$ flow, a plateau representing the $k^{-1}$ power law is observed in the range of
$0.1\leq k\eta\leq 0.3$ (gray region), and the crossover region occurs in $0.06<k\eta<0.1$. However, the $k^{-1}$ power
law disappears when $Sc$ is reduced to $5$. This reveals that in the viscous-convective range, the scalar spectrum in
the $Sc\gg 1$ compressible turbulent mixing defers to the $k^{-1}$ scaling given by the Batchelor theory, which was
previously developed for incompressible turbulence.

On the other hand, when the Schmidt number is $Sc\ll 1$, the scalar fluctuations at scales smaller than the OC scale,
$\eta_{OC}=Sc^{-3/4}\eta$, decay strongly, and thus, the scalar spectrum rolls off steeper than the $k^{-5/3}$
power law. If the Reynolds number is sufficiently large, standard arguments suggest that in the range of
$\eta_{OC}^{-1}\ll k\ll \eta^{-1}$, a so-called inertial-diffusive range exists.
The aforementioned BHT theory predicts that in this range the scalar spectrum has the form
\begin{equation}
E_\phi(k) = \big(C_K/3\big)\langle\epsilon_\phi\rangle\langle\epsilon\rangle^{2/3}\chi^{-3}k^{-17/3}.
\end{equation}
Fig.~\ref{fig:fig4}(b) shows the compensated spectra of scalar according to the BHT variables in high $Re$ flows,
as functions of $k\eta$. We observe that in the $Sc=1/25$ flow, a narrow plateau
representing the $k^{-17/3}$ power law arises, where the scale range is roughly $0.3\leq k\eta\leq 0.4$ (gray region),
which corresponds to $3.4\leq k\eta_{OC}\leq 4.5$. The above result demonstrates that in the inertial-diffusive
range, the scalar spectrum in the $Sc\ll 1$ compressible turbulent mixing follows the $k^{-17/3}$ scaling provided by the BHT theory.

\begin{figure}
\centerline{\includegraphics[width=6.5cm]{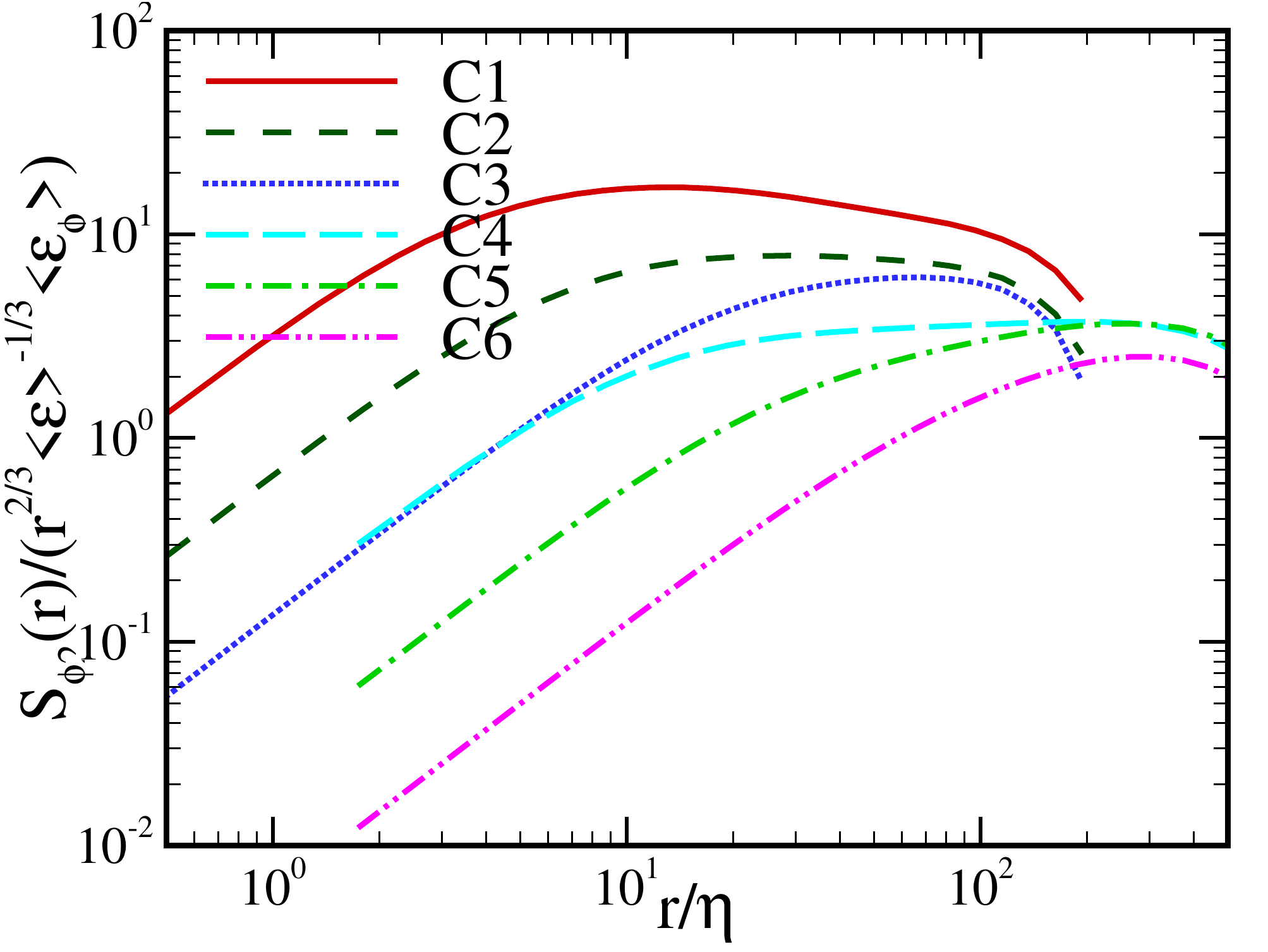}}
\caption{(color online). Obukhov-Corrsin scaling of second-order structure function of scalar at different values of
$Sc$ and $Re_\lambda$.}
\label{fig:fig5}
\end{figure}

The second-order structure function of scalar increment is defined by
\begin{equation}
S_{\phi2}(r)\equiv \langle(\delta_r\phi)^2\rangle,
\end{equation}
where $\delta_r\phi=\phi(\textbf{x}+r)-\phi(\textbf{x})$ is the scalar increment.
In Fig.~\ref{fig:fig5} we plot $S_{\phi2}(r)$ normalized by the OC variables, as suggested in Eq. (3.1), as functions
of the normalized separation distance $r/\eta$. It shows that finite-width plateaus at large $r/\eta$.
The scaling constants computed by the plateaus in low $Re$ flows are higher than those in high $Re$
flows. When $Sc$ decreases, the values of scaling constants in both high and low $Re$ flows become smaller. At sufficiently
large scales, an asymptotic formulation obtained from incompressible turbulent mixing [31] gives
$$\frac{\langle(\delta_r\phi)^2\rangle}{\langle\epsilon_\phi\rangle\langle\epsilon\rangle^{-1/3}r^{2/3}} \approx
\frac{3Re_\lambda Sc^{1/2}}{\sqrt{15}r_\phi}(\frac{r}{\eta_{OC}})^{-2/3}$$
\begin{equation}
=\frac{3Re_\lambda}{\sqrt{15}r_\phi}(\frac{r}{\eta})^{-2/3}.
\end{equation}
Given that the scale range for a plateau is different in every case, in principle it explains the relative relations of
the scaling constants in our simulations.

\begin{figure}
\centerline{\includegraphics[width=6.5cm]{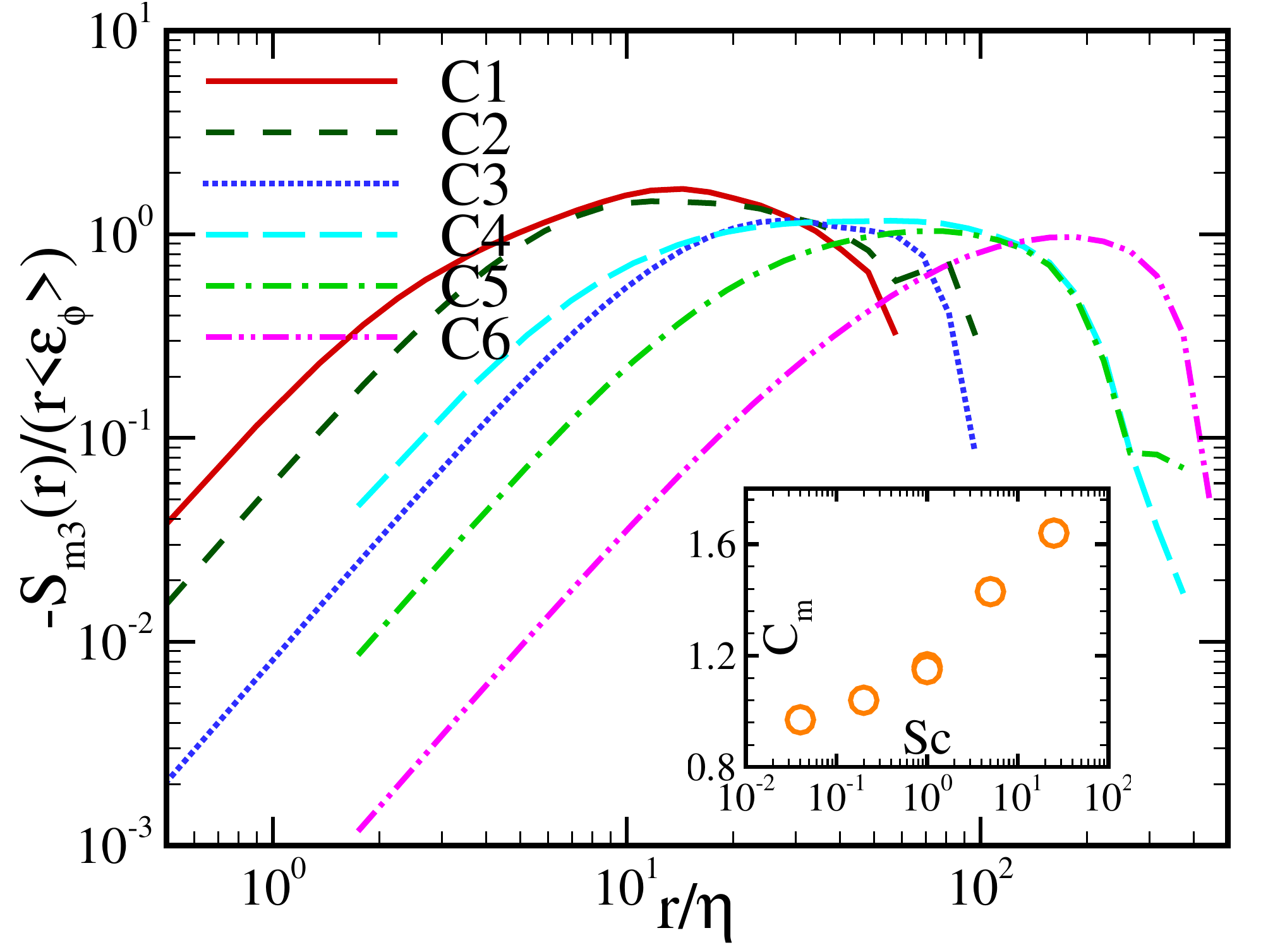}}
\caption{(color online). Yaglom scaling of mixed third-order velocity-scalar structure function at different values of
$Sc$ and $Re_\lambda$.}
\label{fig:fig6}
\end{figure}

The mixed third-order structure function, defined as $S_{m3}(r)=\langle\delta_ru(\delta_r\phi)^2\rangle$, where
$\delta_ru=u(\textbf{x}+r)-u(\textbf{x})$ is the longitudinal velocity increment, plays a more fundamental role in the
similarity scaling. In incompressible turbulence, an exact result for $\eta_B\ll r\ll L_{\phi}$ was given by Yaglom [35]
\begin{equation}
\langle\delta_ru(\delta_r\phi)^2\rangle = -\frac{4}{3}<\epsilon_\phi>r.
\end{equation}
Fig.~\ref{fig:fig6} presents the minus of $S_{m3}(r)$ normalized by the Yaglom variables. It is found that for
each simulated flow, there appears a flat region with finite width. Furthermore, in the limit of large scale, $-S_{m3}(r)$
drops quickly and approaches zero, while at small scales, it behaves approximately as $r^2$ according to the Taylor
expansion. We now employ quantity $C_m$ to represent the compensated mixed third-order structure function
\begin{equation}
C_m = -\frac{S_{m3}}{r\langle\epsilon_\phi\rangle}.
\end{equation}
Our results show that in flat regions, the scaling constant $C_m$ is $1.64$, $1.43$, $1.15$, $1.16$, $1.04$, and
$0.97$ from C1 through C6. Two aspects can thus be concluded: (1) the contribution from the variation in the
Reynolds number to $C_m$ is negligible, while the compressible effect makes $C_m$ smaller than the $4/3$ value from
incompressible turbulence; and (2) $C_m$ has a tendency to increase with $Sc$, which is in agreement with the results
shown in [31]. In the inset of Fig.~\ref{fig:fig6} we plot $C_m$ as a function of $Sc$.

\subsection{\emph{Probability Distribution Function}}

\begin{figure}
\centerline{\includegraphics[width=6.5cm]{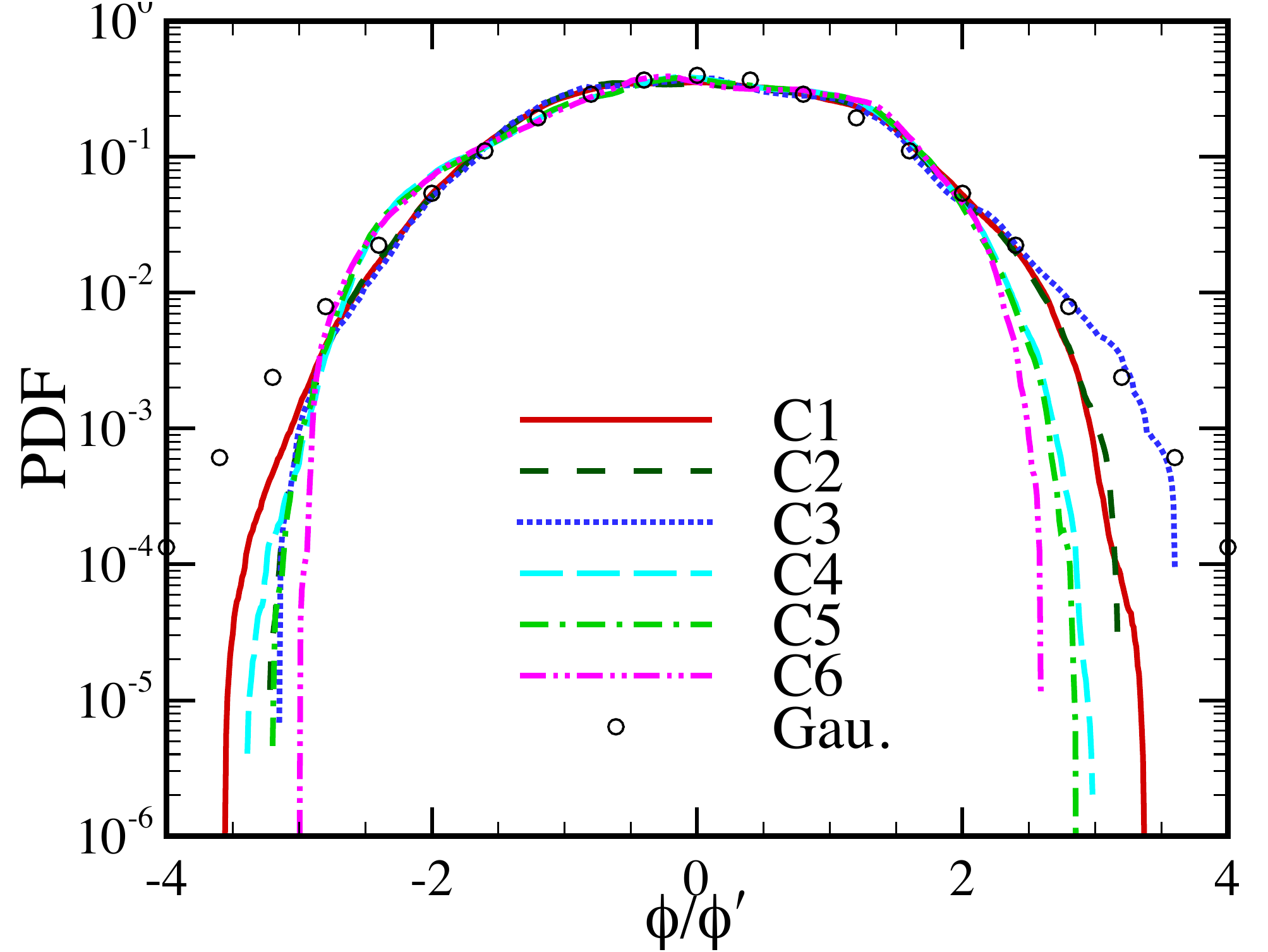}}
\caption{(color online). The one-point PDF of the normalized scalar fluctuations at different values of $Sc$ and
$Re_\lambda$, where the circles are for the Gaussian PDF.}
\label{fig:fig7}
\end{figure}

Fig.~\ref{fig:fig7} shows the one-point PDF of the normalized scalar fluctuations. At small amplitudes the PDFs collapse
to the Gaussian distribution, whereas at large amplitudes they decay more quickly than Gaussian and thus are known as sub-Gaussian.
This feature corresponds to the passive scalar transport in 3D compressible and 1D incompressible turbulent
flows [32,36], which is due to the fact that in our simulations, the ratio of the integral length scale of scalar $L_\phi$
to the computational domain $L_0$ is $0.19\sim 0.21$, which prevents large scalar fluctuations.

\begin{figure}
\centerline{\includegraphics[width=6.5cm]{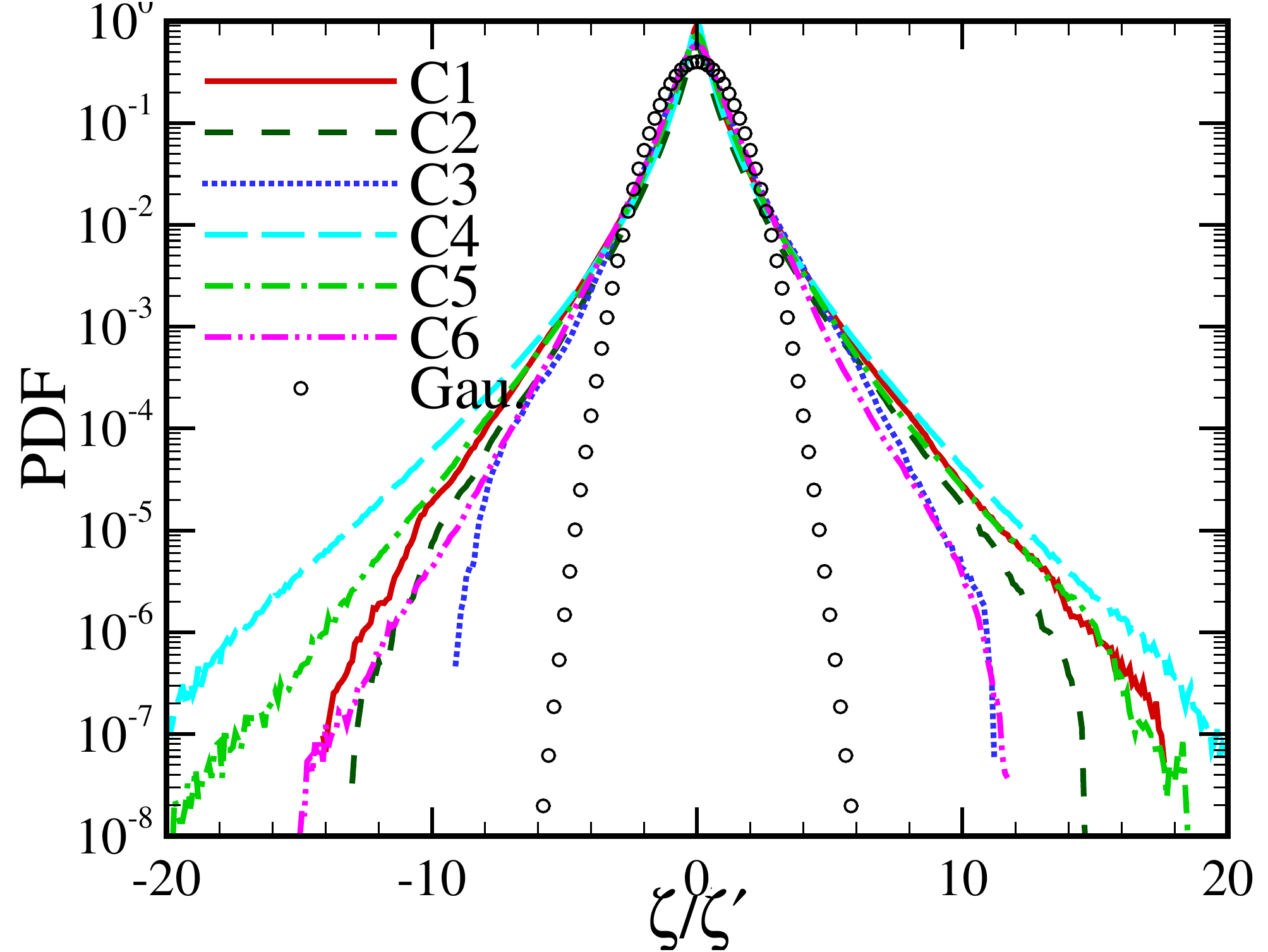}}
\caption{(color online). The one-point PDF of the normalized scalar gradient, where the circles are for the Gaussian PDF.}
\label{fig:fig8}
\end{figure}

In Fig.~\ref{fig:fig8} we plot the one-point PDF of the normalized scalar gradient, where
$\zeta'=\sqrt{\langle(\partial\phi/\partial x_j)^2\rangle}$ is the r.m.s. magnitude of scalar gradient.
Obviously, the convex PDF tails are much longer than those of Gaussian, indicating
strong intermittency. In both high and low $Re$ flows, the PDF tails on each side become broader as $Sc$ increases.
Moreover, in the $Sc=1$ flows, the notable increase in $Re_\lambda$ leads the PDF tails to be significantly wide.
These observations imply that the growth in the Reynolds and Schmidt numbers will reinforce the events of
extreme scalar oscillations at small scales.

\begin{figure}
\begin{center}
\subfigure{
\resizebox*{6.5cm}{!}{\rotatebox{0}{\includegraphics{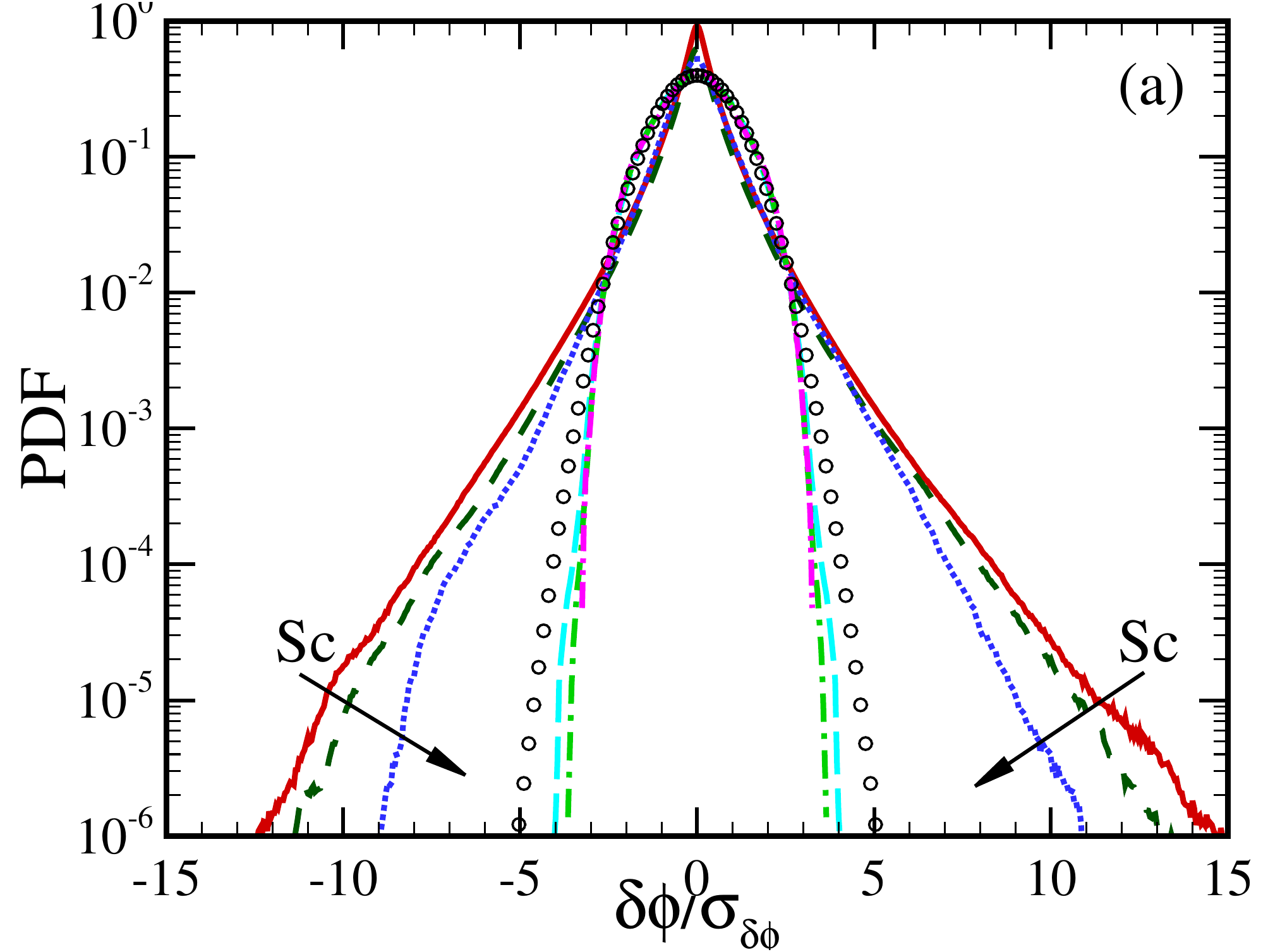}}}}%

\subfigure{
\resizebox*{6.5cm}{!}{\rotatebox{0}{\includegraphics{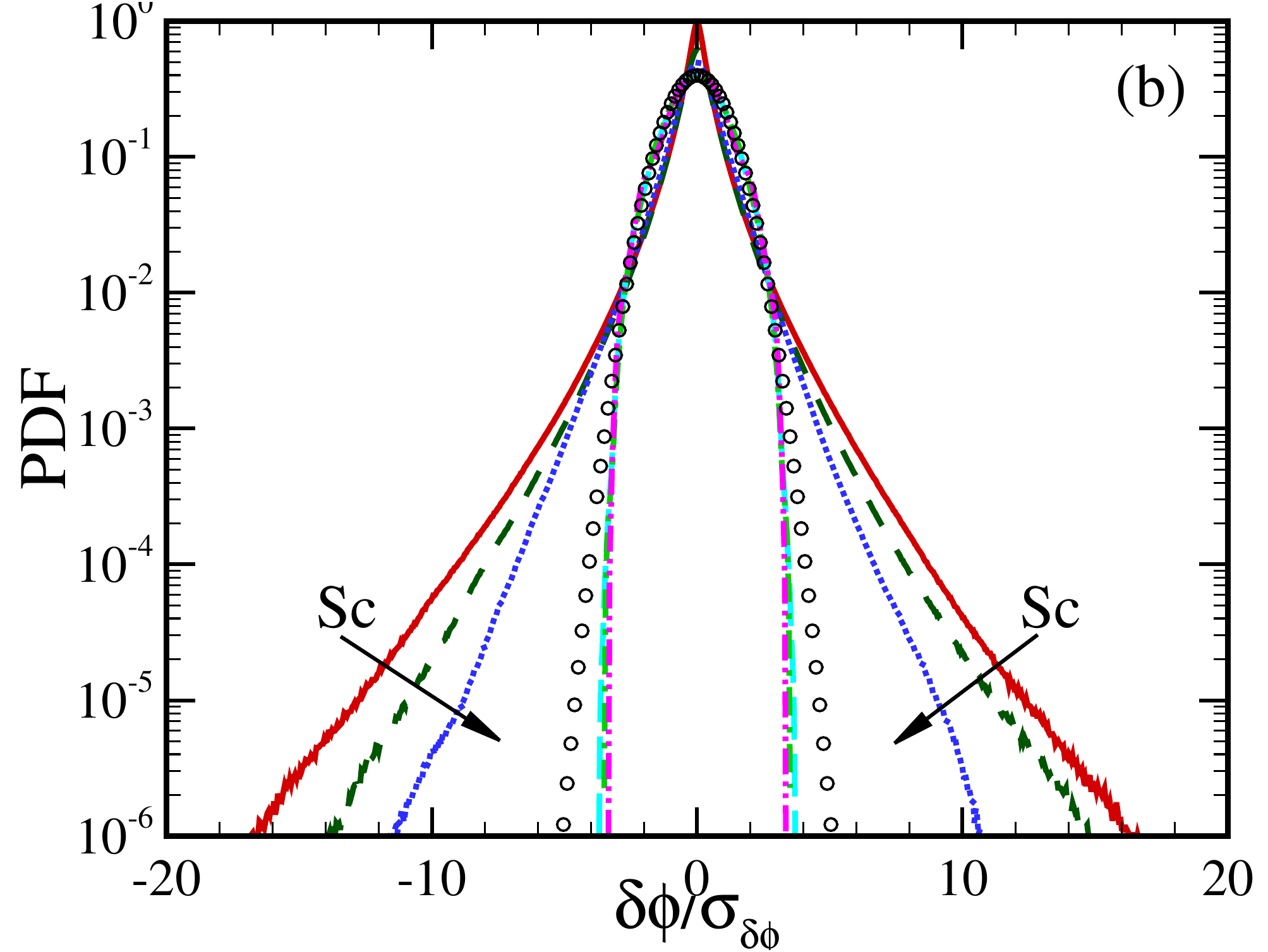}}}}%
\caption{(color online). The two-point PDF of the normalized scalar increment.
($Sc$,$r/\eta$) lines: ($25$,$1$), solid; ($5$,$1$), dashed; ($1$,$1$), dotted;
($25$,$256$), long dashed; ($5$,$256$), dash-dotted; and ($1$,$256$), dash-dot-dotted.
The circles are for the Gaussian PDF, and the arrows indicate the decreasing $Sc$.
(a) low Reynolds number, (b) high Reynolds number.}
\label{fig:fig9}
\end{center}
\end{figure}

As a further study, in Figs.~\ref{fig:fig9}(a) and \ref{fig:fig9}(b) we plot the two-point PDFs of scalar increment against
the normalized separation distances of $r/\eta=1$ and $256$, where $\sigma_{\delta\phi}$ is
the standard deviation of $\delta\phi$. We find that in both high and low $Re_\lambda$ flows, the behaviors of the PDF tails
at large and small scales are respectively similar to the one-point PDF tails shown in Figs.~\ref{fig:fig7} and \ref{fig:fig8}.

\begin{figure}
\centerline{\includegraphics[width=6.5cm]{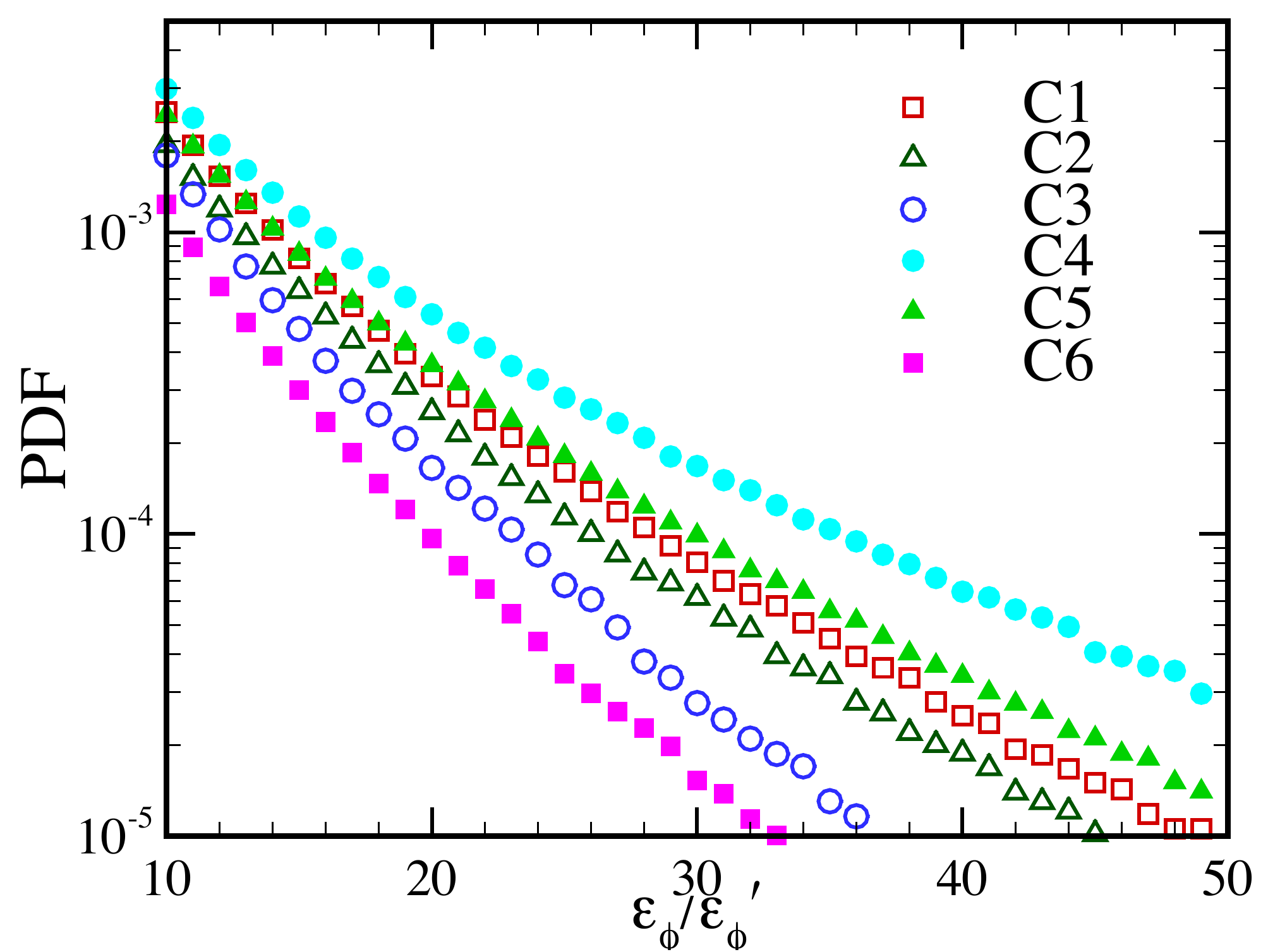}}
\caption{(color online). The PDF of the normalized scalar dissipation rate at different values of $Sc$ and $Re_\lambda$.}
\label{fig:fig10}
\end{figure}

Many studies in the literature of incompressible turbulent flows suggest that, at small scales, the intermittency of scalar field is closely
associated with its dissipation. Fig.~\ref{fig:fig10} shows the PDFs of the normalized scalar dissipation rate, where $\epsilon_{\phi}'=\sqrt{\langle(\epsilon_{\phi}-\langle\epsilon_{\phi}\rangle)^2\rangle}$
is the r.m.s. magnitude of $\epsilon_\phi$. Similar to that shown in Fig.~\ref{fig:fig8}, it is found that an increase in the Schmidt
number strengthens the scalar dissipation occurring at large magnitudes.

\subsection{\emph{Scalar Transport Analysis: high Sc versus low Sc}}

\begin{figure}
\begin{center}
\subfigure{
\resizebox*{6.5cm}{!}{\rotatebox{-90}{\includegraphics{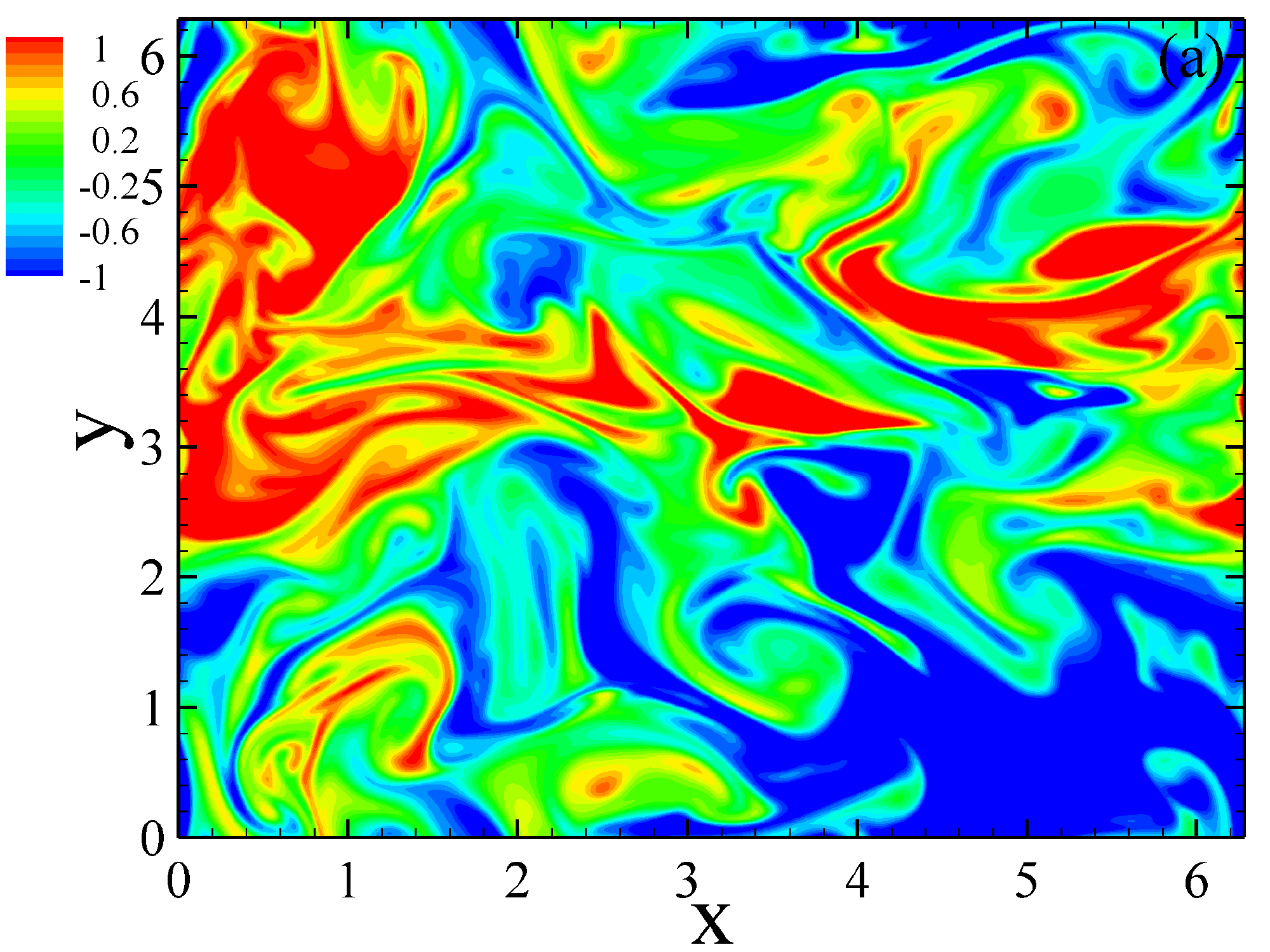}}}}%

\subfigure{
\resizebox*{6.5cm}{!}{\rotatebox{-90}{\includegraphics{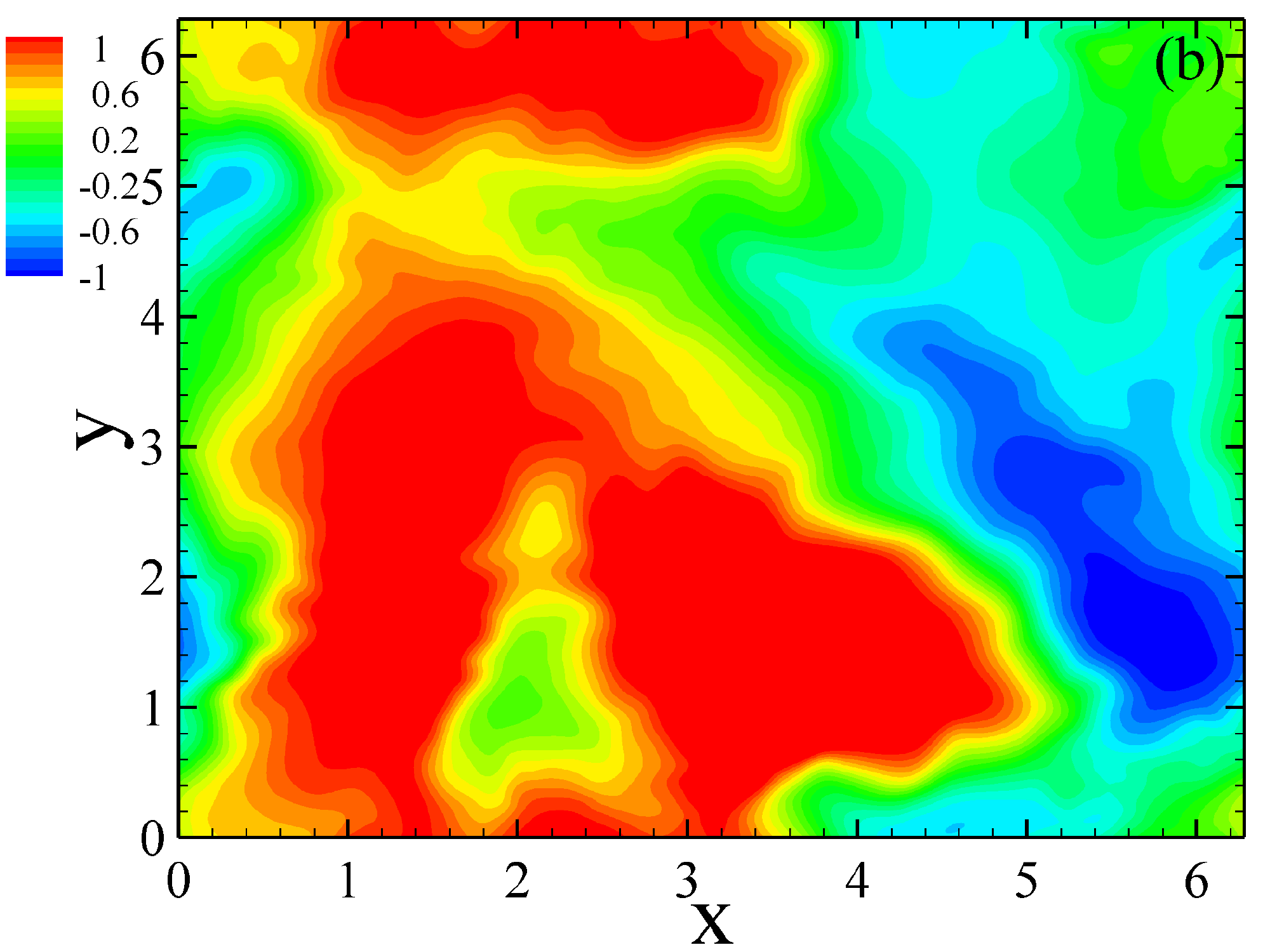}}}}%
\caption{(color online). Two-dimensional contours of scalar field in (a) C1 and (b) C6, at $z=\pi/2$.}
\label{fig:fig11}
\end{center}
\end{figure}

A central problem in turbulent mixing is the transport of scalar fluctuations in the inertial-convective range. For
$Sc\gg 1$ or $Sc\ll 1$, this process should be also related to the viscous-convective or inertial-diffusive
range. Fig.~\ref{fig:fig11} shows the two-dimensional contours of scalar fields in the $z=\pi/2$ plane for
the $Sc=25$ and $1/25$ flows. In the highest $Sc$ flow
the low molecular diffusivity leads the scalar field to roll up and sufficiently mix. Nevertheless, in the lowest $Sc$ flow
the scalar field loses the small-scale structures by the high molecular diffusivity and leaves the large-scale,
cloudlike structures. Given that there are density fluctuations in compressible turbulence, we introduce the density weighted
scalar $\Phi=\sqrt{\rho}\phi$. The governing equation of scalar variance is then obtained by Eqs.(2.1) and (2.4) as follows
$$\frac{\partial}{\partial t}\big(\frac{\Phi^2}{2}\big)
= - u_j\frac{\partial}{\partial x_j}\big(\frac{\Phi^2}{2}\big)
- 2\theta\big(\frac{\Phi^2}{2}\big) -\frac{\chi}{\beta}\big(\frac{\partial\phi}{\partial x_j}\big)^2$$
\begin{equation}
+ \frac{\chi}{\beta}\frac{1}{\sqrt{\rho}}\frac{\partial^2}{\partial x_j^2}\big(\frac{\Phi^2}{2}\big).
\end{equation}
Here, the terms on the right-hand-side of Eq. (3.9) successively stand for the advection, scalar-dilatation, dissipation and
diffusion. In statistically homogeneous turbulence, the global average on the diffusion makes it vanish.
Therefore, in a statistically stationary state, we only focus the first three terms.

\begin{figure}
\begin{center}
\subfigure{
\resizebox*{6.5cm}{!}{\rotatebox{-90}{\includegraphics{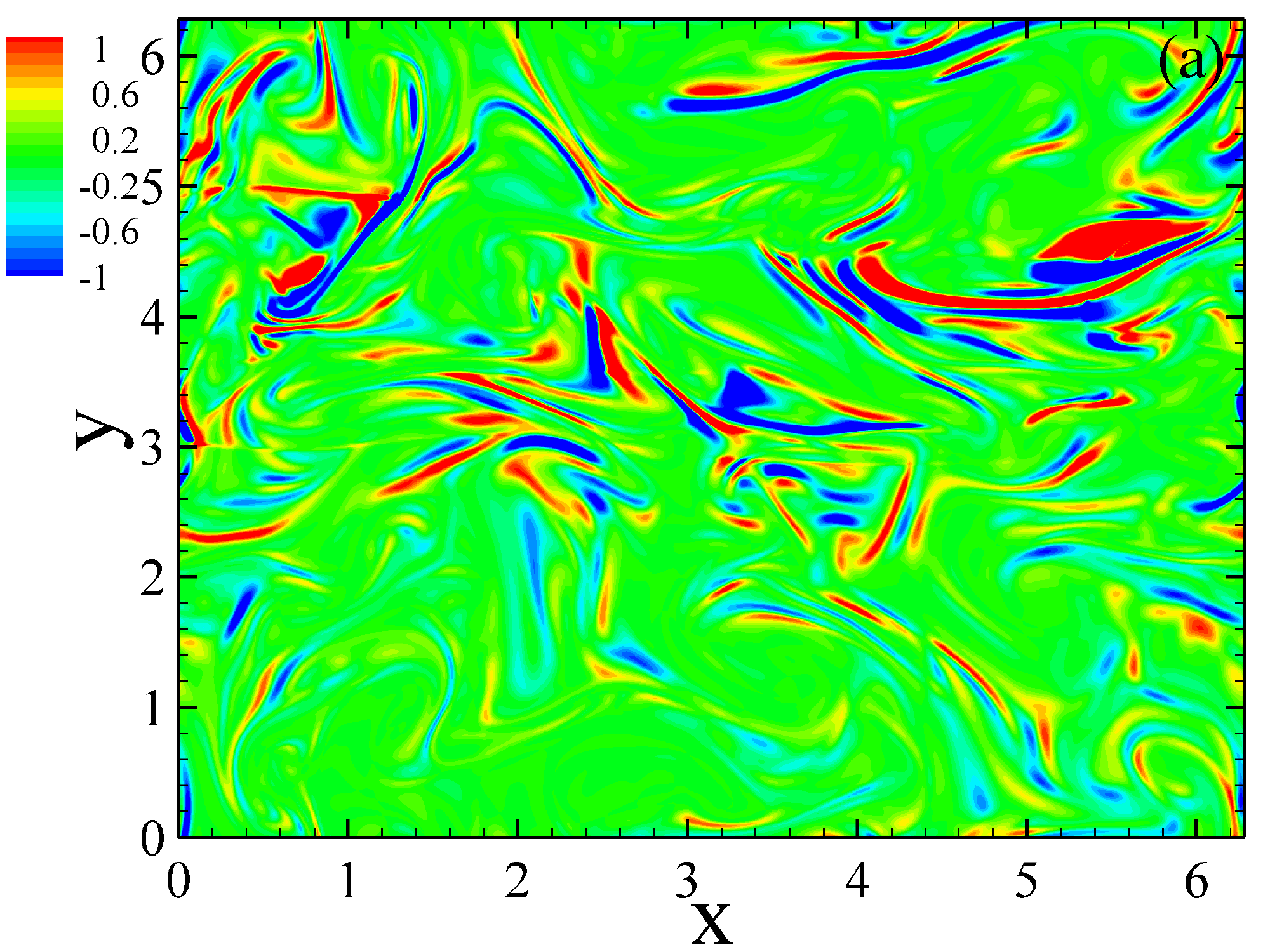}}}}%

\subfigure{
\resizebox*{6.5cm}{!}{\rotatebox{-90}{\includegraphics{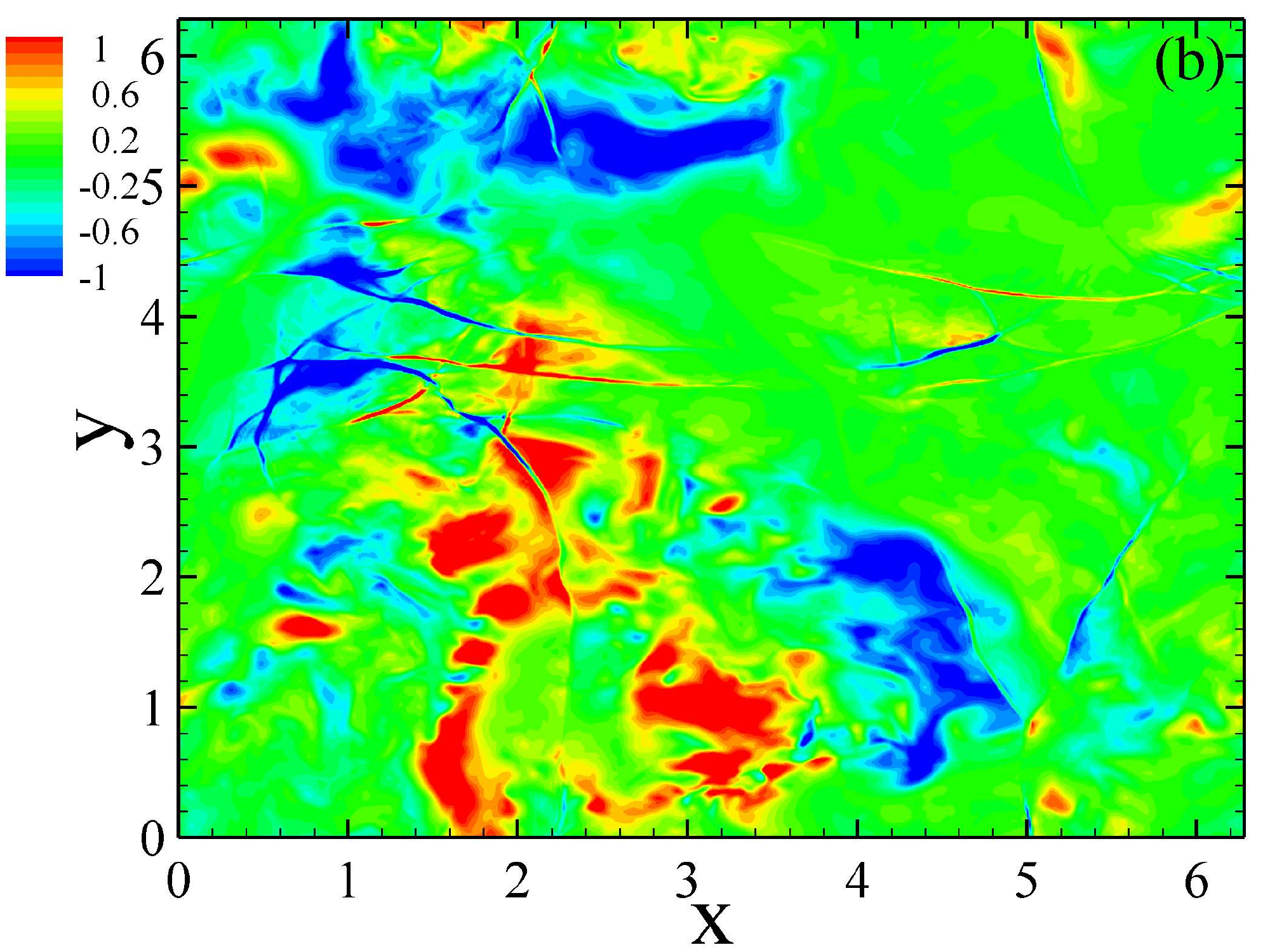}}}}%
\caption{(color online). Two-dimensional contours of the advection term of the scalar variance equation
in (a) C1 and (b) C6, at $z=\pi/2$.}
\label{fig:fig12}
\end{center}
\end{figure}

\begin{figure}
\begin{center}
\subfigure{
\resizebox*{6.5cm}{!}{\rotatebox{-90}{\includegraphics{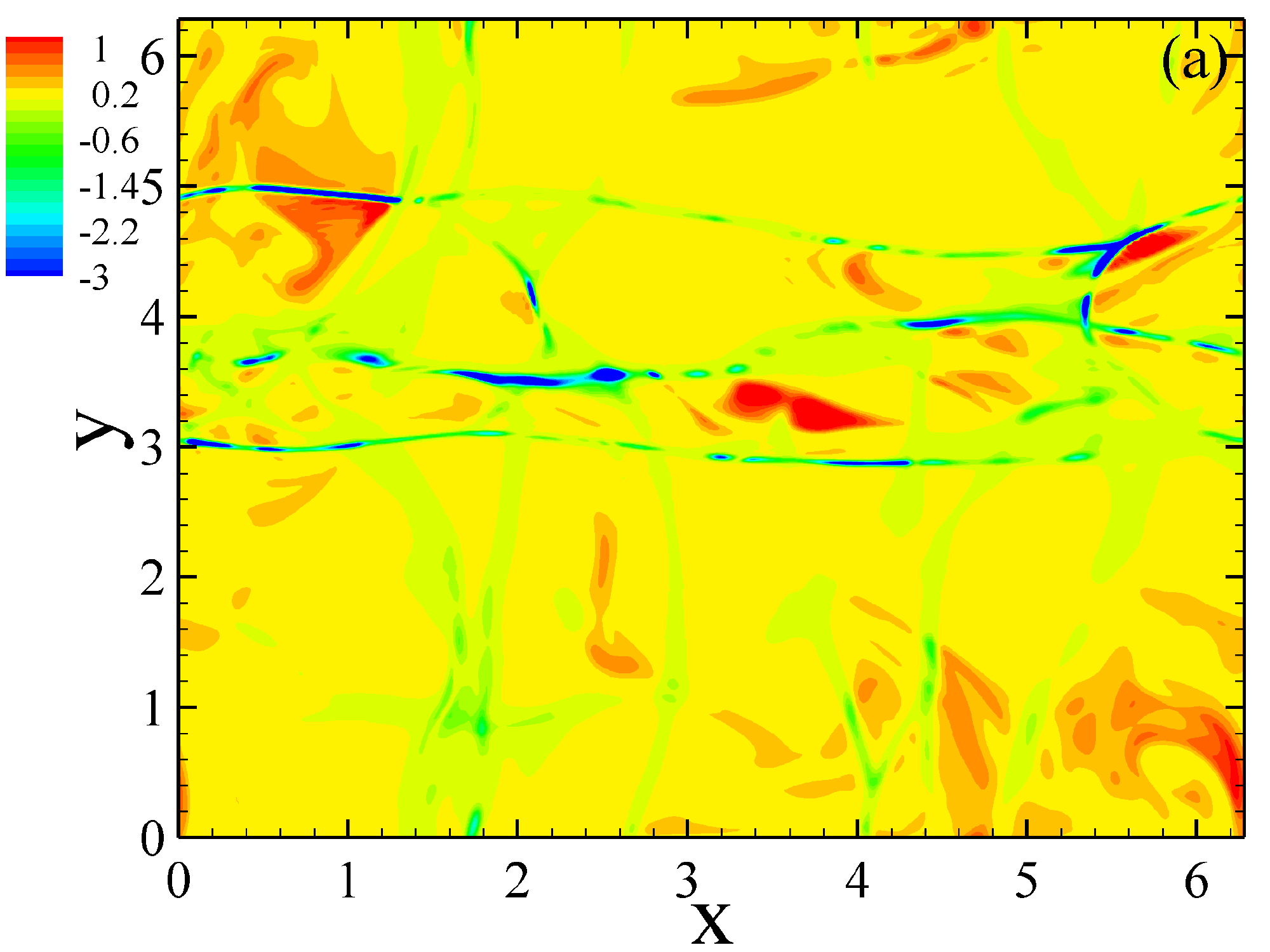}}}}%

\subfigure{
\resizebox*{6.5cm}{!}{\rotatebox{-90}{\includegraphics{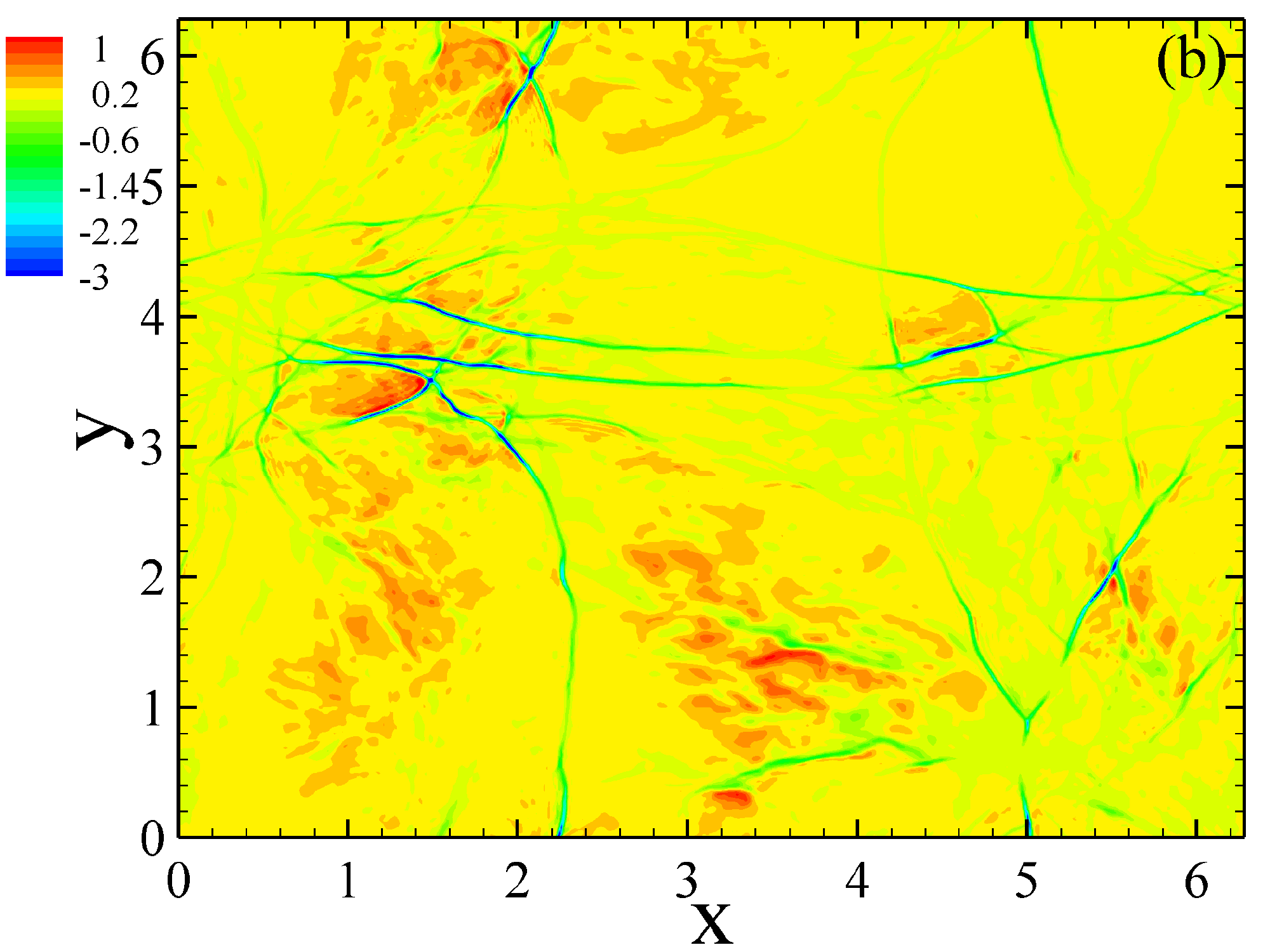}}}}%
\caption{(color online). Two-dimensional contours of the scalar-dilatation term of the scalar
variance equation in (a) C1 and (b) C6, at $z=\pi/2$.}
\label{fig:fig13}
\end{center}
\end{figure}

\begin{figure}
\begin{center}
\subfigure{
\resizebox*{6.5cm}{!}{\rotatebox{-90}{\includegraphics{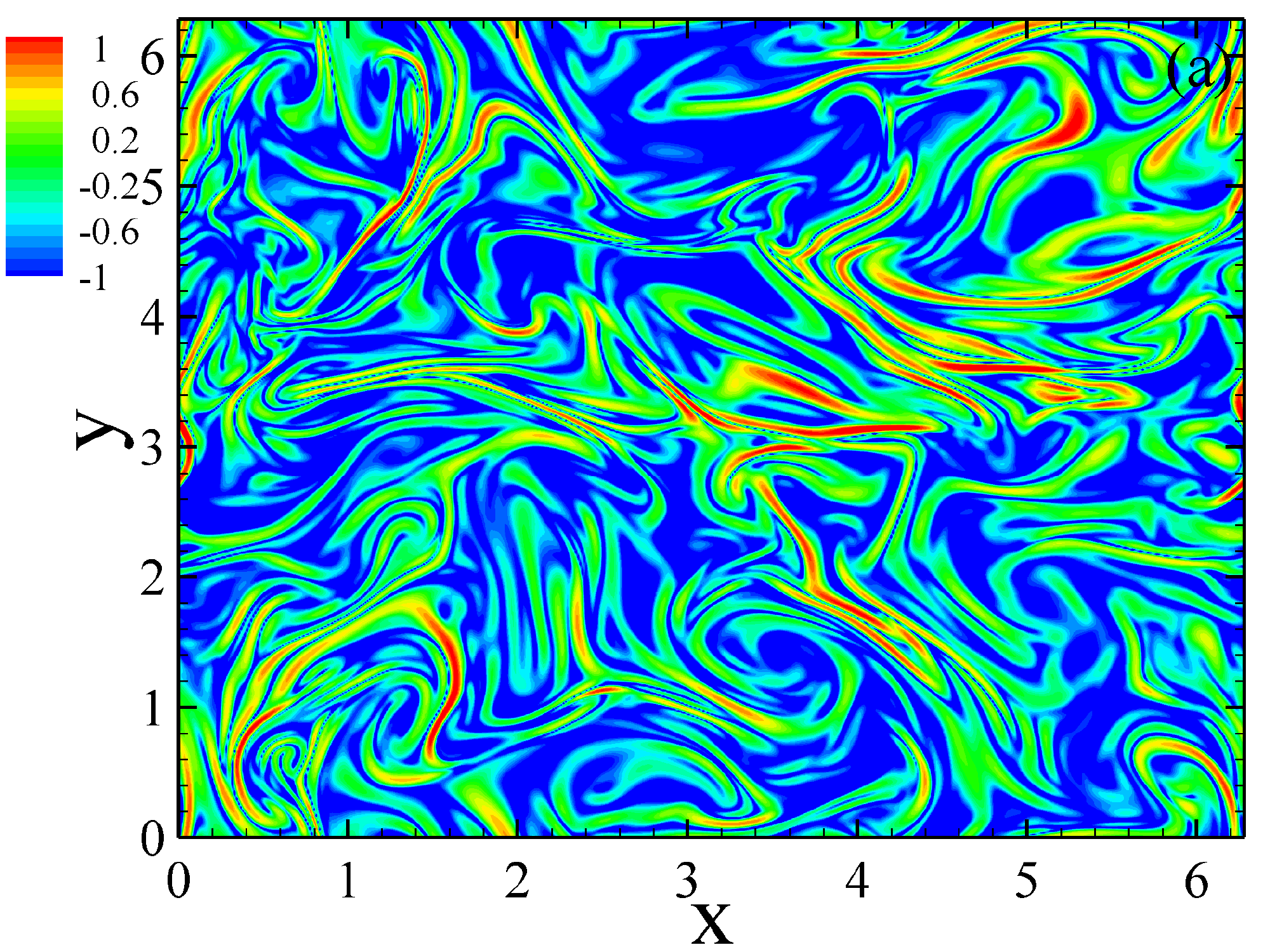}}}}%

\subfigure{
\resizebox*{6.5cm}{!}{\rotatebox{-90}{\includegraphics{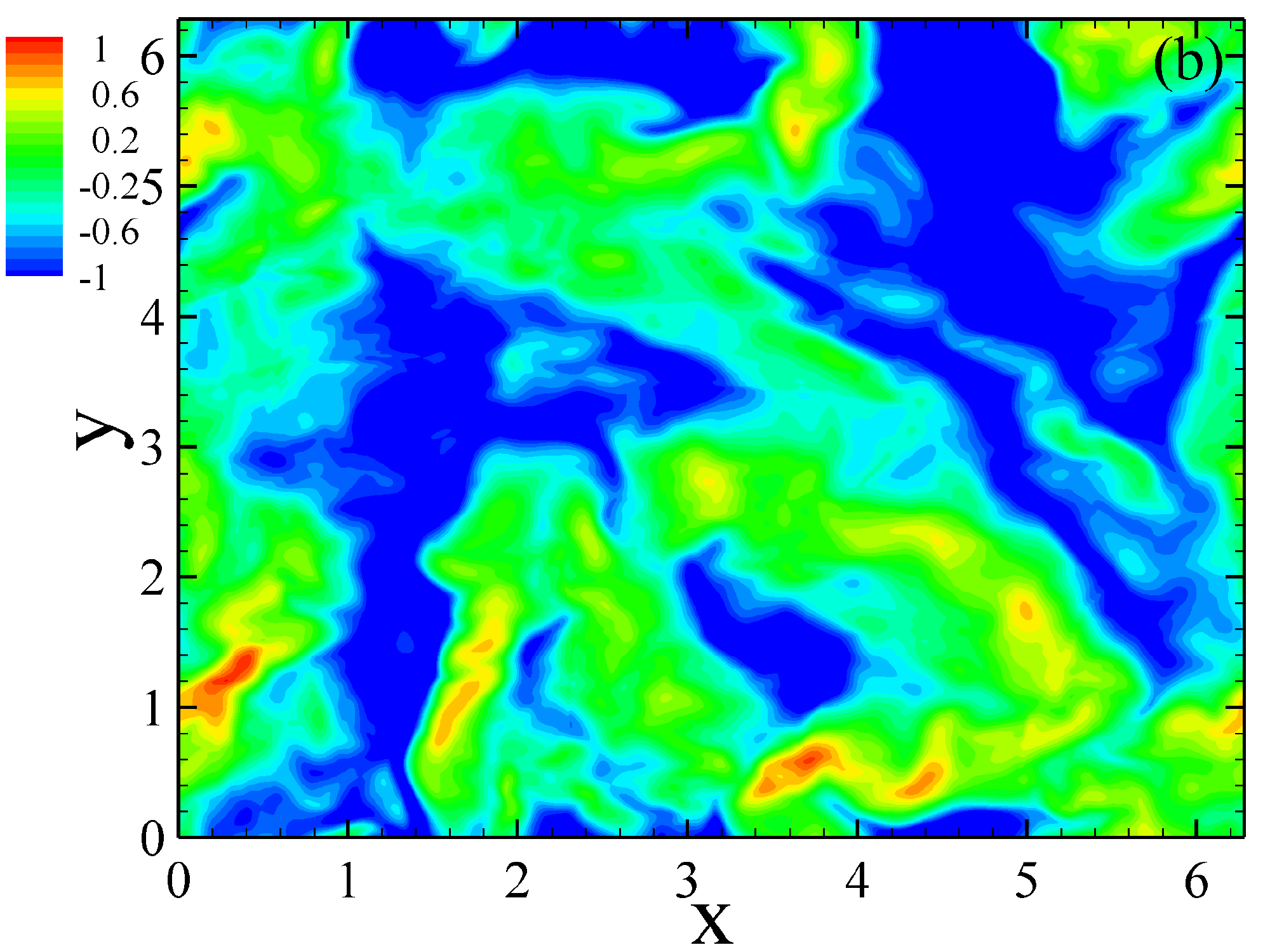}}}}%
\caption{(color online). Two-dimensional contours of logarithm of the dissipation term of the scalar
variance equation in (a) C1 and (b) C6, at $z=\pi/2$.}
\label{fig:fig14}
\end{center}
\end{figure}

In Fig.~\ref{fig:fig12} we present the two-dimensional contours of the advection terms in the $z=\pi/2$ plane for
the $Sc=25$ and $1/25$ flows. In the highest $Sc$ flow, the small-scale regions of extreme scalar advection
distribute approximately randomly in space, and display as
thin streamers. For the lowest $Sc$ flow, the small-scale structures are basically smeared by the high molecular
diffusivity, and thus, the remaining structures are large in scale and exhibited as clouds. Fig.~\ref{fig:fig13}
shows the contours of the scalar-dilatation terms in the $z=\pi/2$ plane for the same flows. Because of the strong
degree of compressibility induced by forcing, there appear
large-scale shock waves in the scalar-dilatation contours. Besides, in the vicinity of a shock front, the scalar undergoes
drastic changes. To observe the detailed structures at both small and large amplitudes, in Fig.~\ref{fig:fig14}
we plot the logarithms of the dissipation term in the $z=\pi/2$ plane, where the color scale is determined
as follows
\begin{equation}
\psi=\log_{10}\big(D/D'\big).
\end{equation}
Here, $D=\chi\big(\partial\phi/\partial x_j\big)^2/\beta$, and $D'=\sqrt{\langle(D-\langle D\rangle)^2\rangle}$
is the r.m.s. magnitude of $D$. The color changes from blue to red when dissipation increases. In the highest $Sc$ flow,
the contour shows that the extreme
dissipation regions are sufficiently mixed and randomly distributed. In contrast, the high molecular diffusivity in the
lowest $Sc$ flow leads the contour to retain only the large-scale cloudlike structures.

\begin{figure}
\begin{center}
\subfigure{
\resizebox*{6.5cm}{!}{\rotatebox{0}{\includegraphics{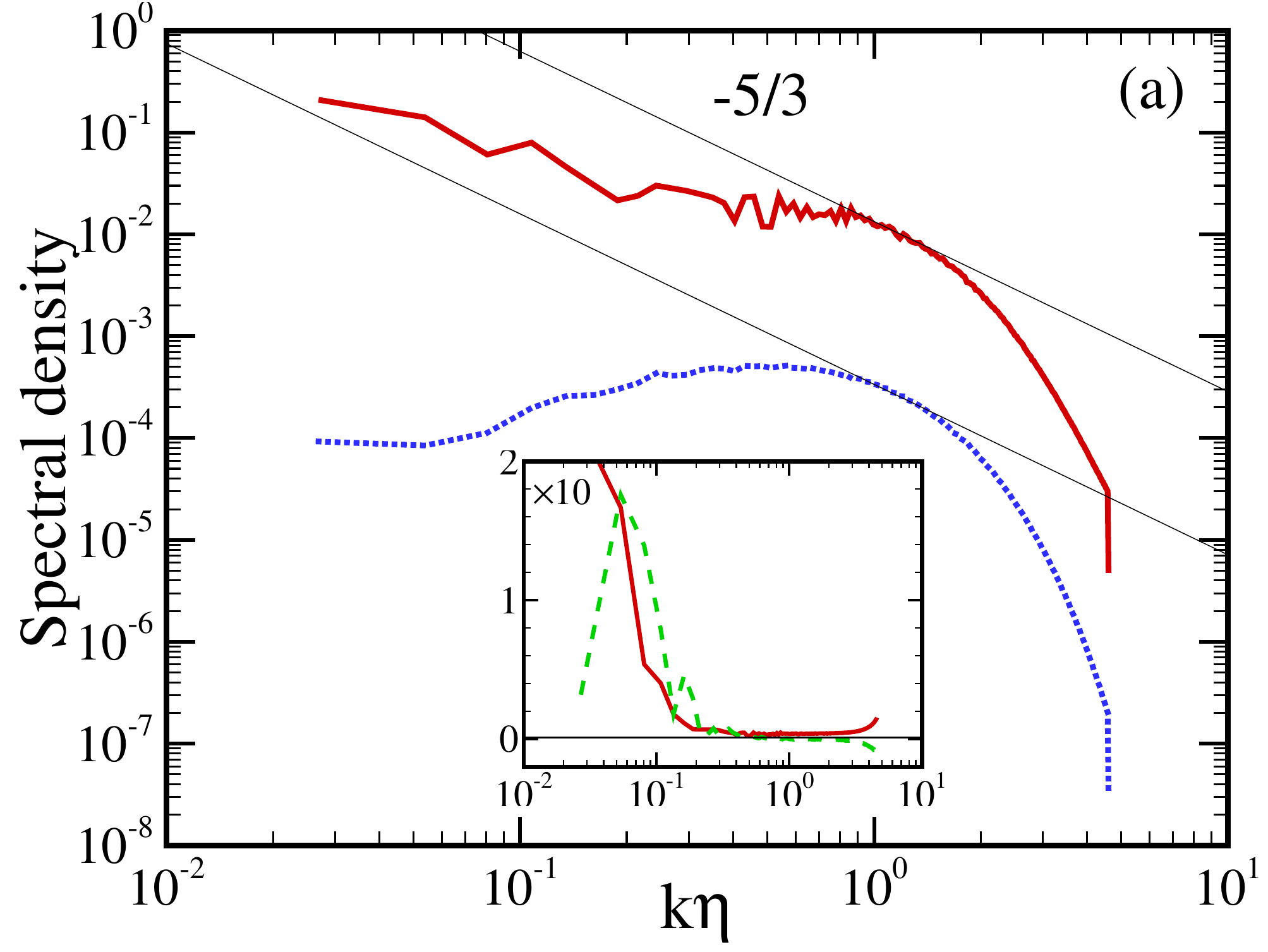}}}}%

\subfigure{
\resizebox*{6.5cm}{!}{\rotatebox{0}{\includegraphics{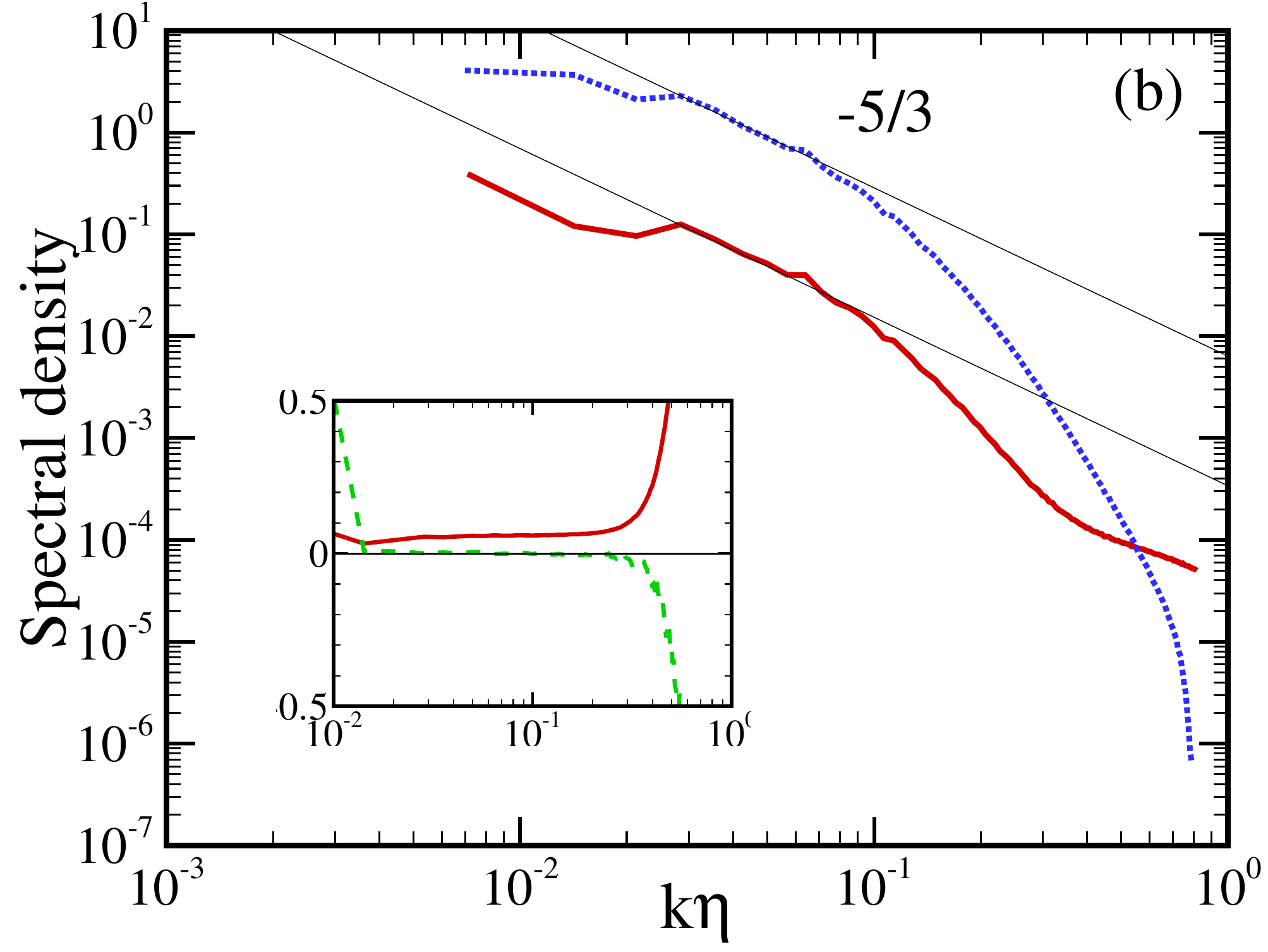}}}}%
\caption{(color online). Spectral densities of terms in the scalar variance equation in Fourier space, where the solid and
dotted lines are for the advection and dissipation terms, respectively. Inset: advection (solid line) and scalar-dilatation
(dashed line) spectra normalized by the dissipation spectrum. (a) C1, (b) C6.}
\label{fig:fig15}
\end{center}
\end{figure}

We now take attention to the spectral analysis of the transport of scalar fluctuations. First, Eq.(3.9) in Fourier space
is written as follows
$$\hat{\Phi}^*(\textbf{k})\frac{\partial}{\partial t}\hat{\Phi}(\textbf{k})
= -\hat{\Phi}^*(\textbf{k})\widehat{\textbf{u}\cdot\nabla\Phi}(\textbf{k})
-\hat{\Phi}^*(\textbf{k})\widehat{\theta\Phi}(\textbf{k})$$
\begin{equation}
+\frac{\chi}{\beta}\hat{\Phi}^*(\textbf{k})\widehat{\frac{1}{\sqrt{\rho}}\nabla^2\Phi}(\textbf{k}),
\end{equation}
where the carets denote the Fourier coefficients, and the asterisks denote complex conjugates. In Fig.~\ref{fig:fig15} we depict
the log-log plots of the spectral densities of the advection and dissipation terms from Eq.(3.11). It is observed that in $Sc=25$
and $1/25$ flows, there appear $k^{-5/3}$ power laws for both advection and dissipation. This indicates that although in compressible
turbulent mixing, the presence of large-scale shock waves significantly affects the transport of scalar fluctuations,
the processes of advection and dissipation may follow the Kolmogorov picture. {Note that the process of scalar-dilatation coupling
may not obey the Kolmogorov picture}. The insets show the spectra of
advection and scalar-dilatation normalized by the dissipation spectrum. In the highest $Sc$ flow, both the advection and
scalar-dilatation spectra fall quickly as wavenumber increases. In contrast, in the lowest $Sc$ flow the advection spectrum
increases at large wavenumbers; however, the scalar-dilatation spectrum is positive at small wavenumbers but becomes negative
at large wavenumbers, indicating mutually opposing scalar transfer processes.

\begin{figure}
\centerline{\includegraphics[width=6.5cm]{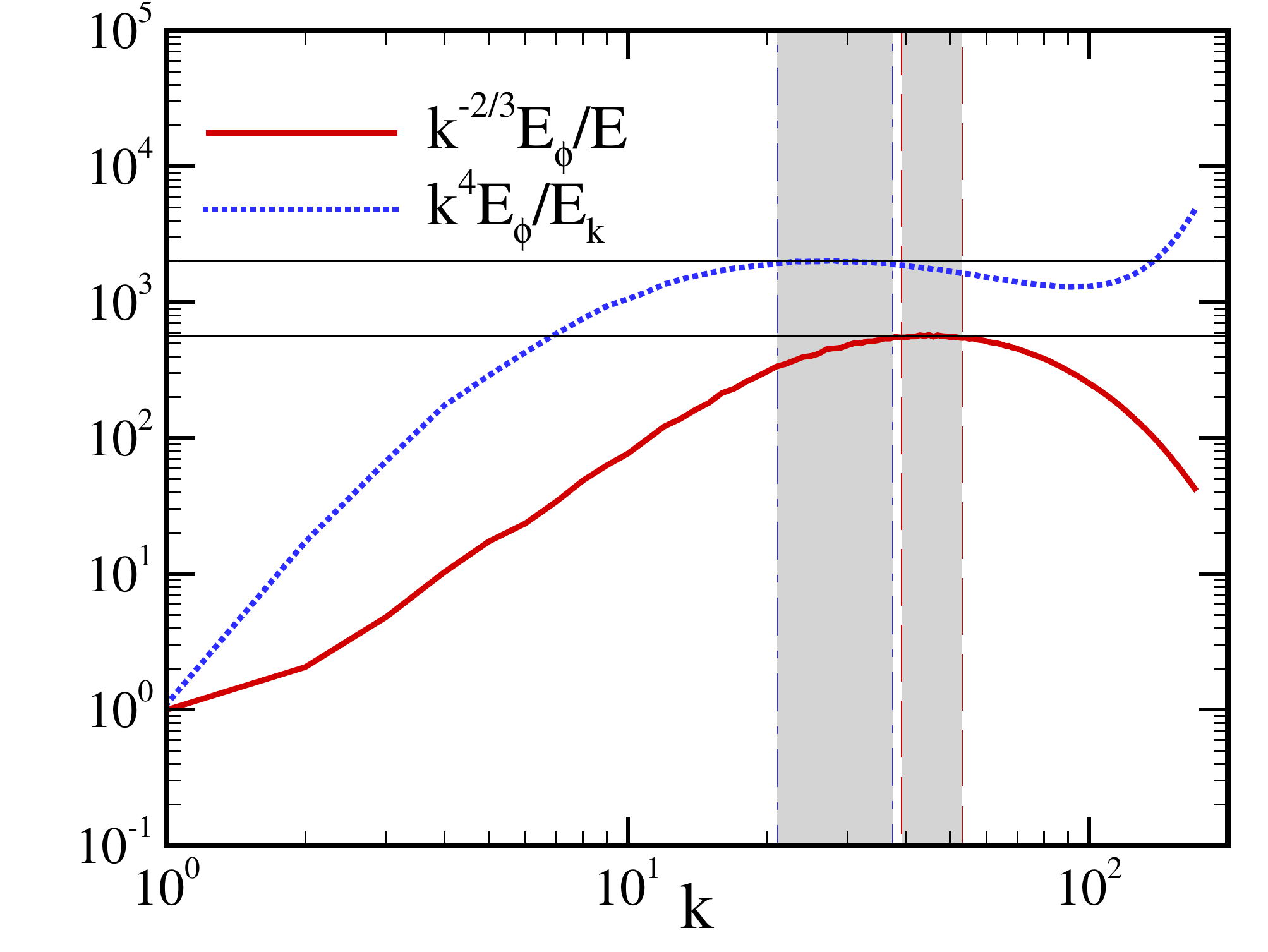}}
\caption{(color online). The relative compensated scalar spectra of $k^{-2/3}E_{\phi}/E_k$ and $k^{4}E_{\phi}/E_k$ in
C1 and C6, respectively.}
\label{fig:fig16}
\end{figure}

To determine the scalar spectra in both highest and lowest $Sc$ flows, we compute the compensated scalar spectra
relative to the kinetic energy spectrum. According to the aforementioned theories, the results are $k^{-2/3}E_\phi/E_k$ at
$Sc=25$ and $k^4E_\phi/E_k$ at $Sc=1/25$. It shows that plateaus appear for $k^{-2/3}E_\phi/E_k$ in the
range of {$38\leq k\leq 51$ (gray region with dashed edge lines)} and $k^4E_\phi/E_k$
in the range of {$23\leq k\leq 34$ (gray region with dash-dotted edge lines)}, respectively, which
indicates that the related spectra defined in the viscous-convective and inertial-diffusive ranges have flat regions
located at relative large and small wavenumbers, respectively. Undoubtedly, the kinetic energy spectrum for both flows
have the inertial range of $E_k(k)=C_k\langle\epsilon\rangle^{2/3}k^{-5/3}$ [21], where $C_k$ is the
Kolmogorov constant. This in turn yields $E_\phi$ as follows
\begin{equation}
E_\phi \propto C_k\langle\epsilon\rangle^{2/3} k^{-1}, \quad Sc=25;
\end{equation}
\begin{equation}
E_\phi \propto C_k\langle\epsilon\rangle^{2/3} k^{-17/3}, \quad Sc=1/25.
\end{equation}

\begin{figure}
\centerline{\includegraphics[width=6.5cm]{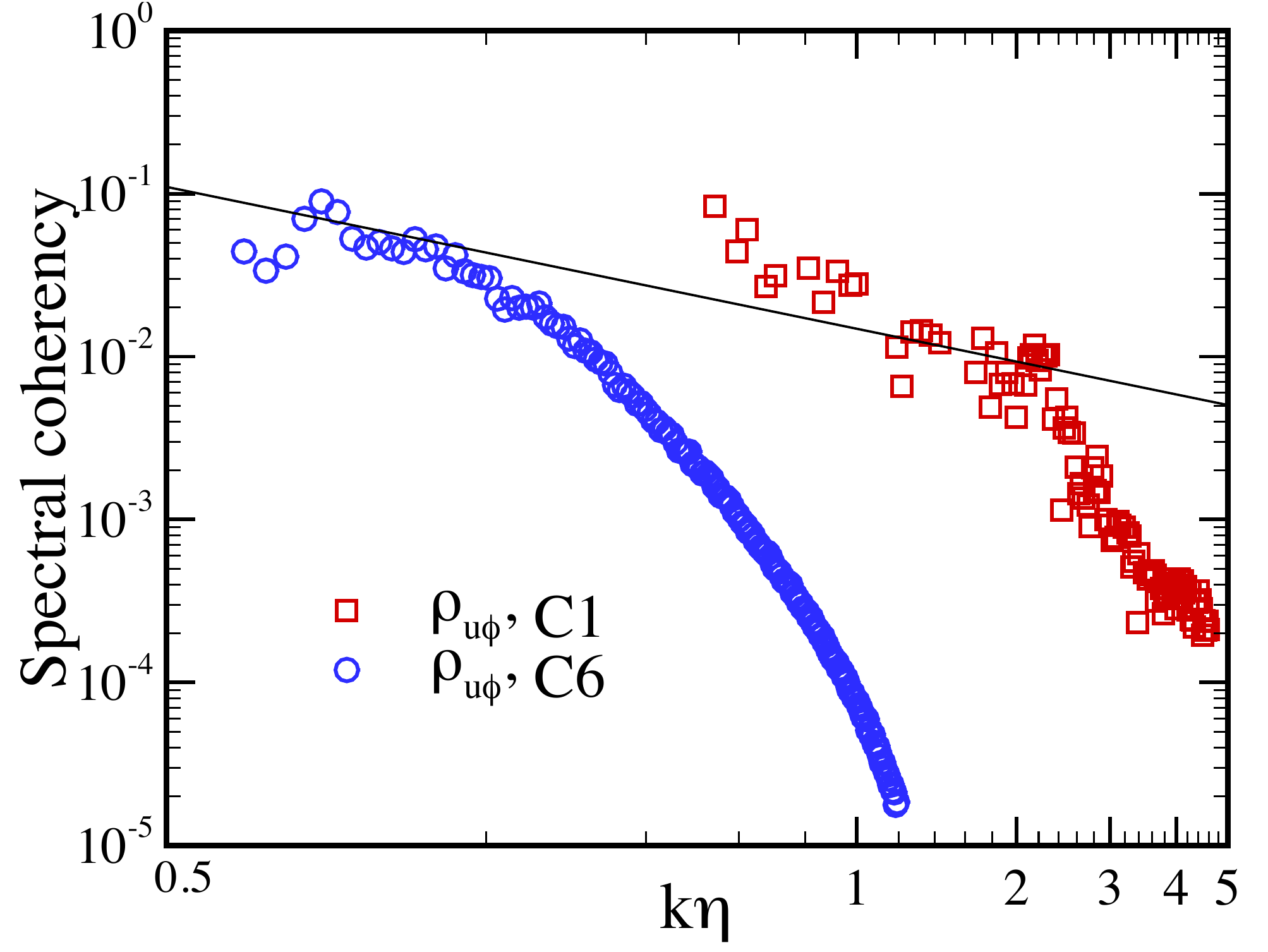}}
\caption{(color online). Spectral coherency of scalars in C1 and C6. The slope value of line is $-2/3$.}
\label{fig:fig17}
\end{figure}

It is useful to work with the spectral coherency defining as
\begin{equation}
\rho_{u\phi}(k)\equiv E_{u\phi}(k)/[E_k(k)E_\phi(k)]^{1/2}.
\end{equation}
Here, $E_{u\phi}(k)$ is the cospectrum of velocity and scalar, which is defined by [37]
\begin{equation}
E_{u\phi}(k)\equiv \int dS_k\langle u(\textbf{k})\phi^*(\textbf{k})\rangle,
\end{equation}
where the integral $\int dS_k$ is taken over a spherical shell in wavenumber space. Lumley [38] proposed
a $k^{-7/3}$ power law for $E_{u\phi}(k)$ in the inertial-convective range, which was in good agreement with
a recent simulation study [37]. Since both $E_k(k)$ and $E_\phi(k)$ scale as $k^{-5/3}$ under similar conditions,
the corresponding result for $\rho_{u\phi}(k)$ must be $k^{-2/3}$. In Fig.~\ref{fig:fig17}, at high wavenumbers,
the spectral coherency in the $Sc=25$ and $1/25$ flows decay faster than $k^{-2/3}$. This deviation is mainly
caused by the contribution from the coupling of scalar and dissipation, which provides
a different power law. Contrary to the observations from incompressible turbulent mixing [15], here when $Sc$ increases,
the spectral coherency falls more slowly with wavenumber. Furthermore, the scattering of the spectral coherency points
in the higher $Sc$ flow is because of the lower Reynolds number.

\subsection{\emph{Comparison Between Compressible and Incompressible Results}}

\begin{figure}
\centerline{\includegraphics[width=6.5cm]{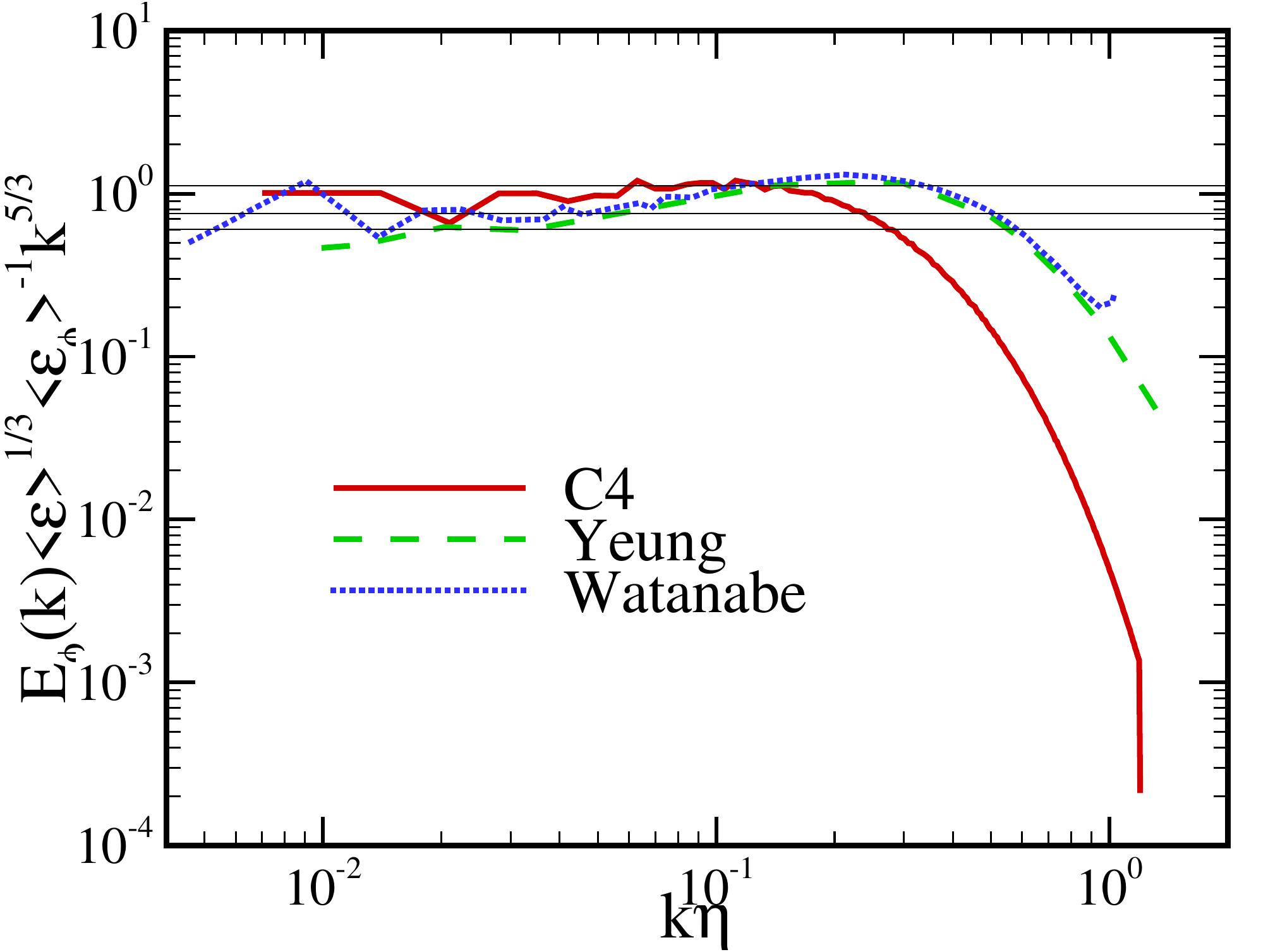}}
\caption{(color online). Compensated spectrum of scalar according to the Obukhov-Corrsin variables. The data from C4, [31] and [32]
are denoted by the solid, dashed and dotted lines, respectively.}
\label{fig:fig18}
\end{figure}

In the final subsection we discuss the comparisons of certain results between compressible and incompressible turbulence.
Herein, we focus solely on cases of Schmidt number at unity. In Fig.~\ref{fig:fig18} we plot the compensated scalar spectra
according to the OC variables from C4, [31] and [32], where the values of $Re_\lambda$ are
$208$, $240$ and $258$, respectively. For C4 the plateau centered at around $k\eta\approx 0.11$ gives that $C_\phi\approx 1.12$.
In contrast, the plateaus from the two incompressible flows are centered at lower wavenumbers, and the values of
$C_\phi$ are $0.67\sim 0.68$. The spectral bumps appearing in the two incompressible flows are quite conspicuous.
However, in the compressible flow it is difficult to identify a clear bump. Furthermore, owing to the contribution
of dissipation from shock waves, in the dissipative range, the scalar spectrum of compressible flow decays more
quickly than its two incompressible counterparts.

\begin{figure}
\centerline{\includegraphics[width=6.5cm]{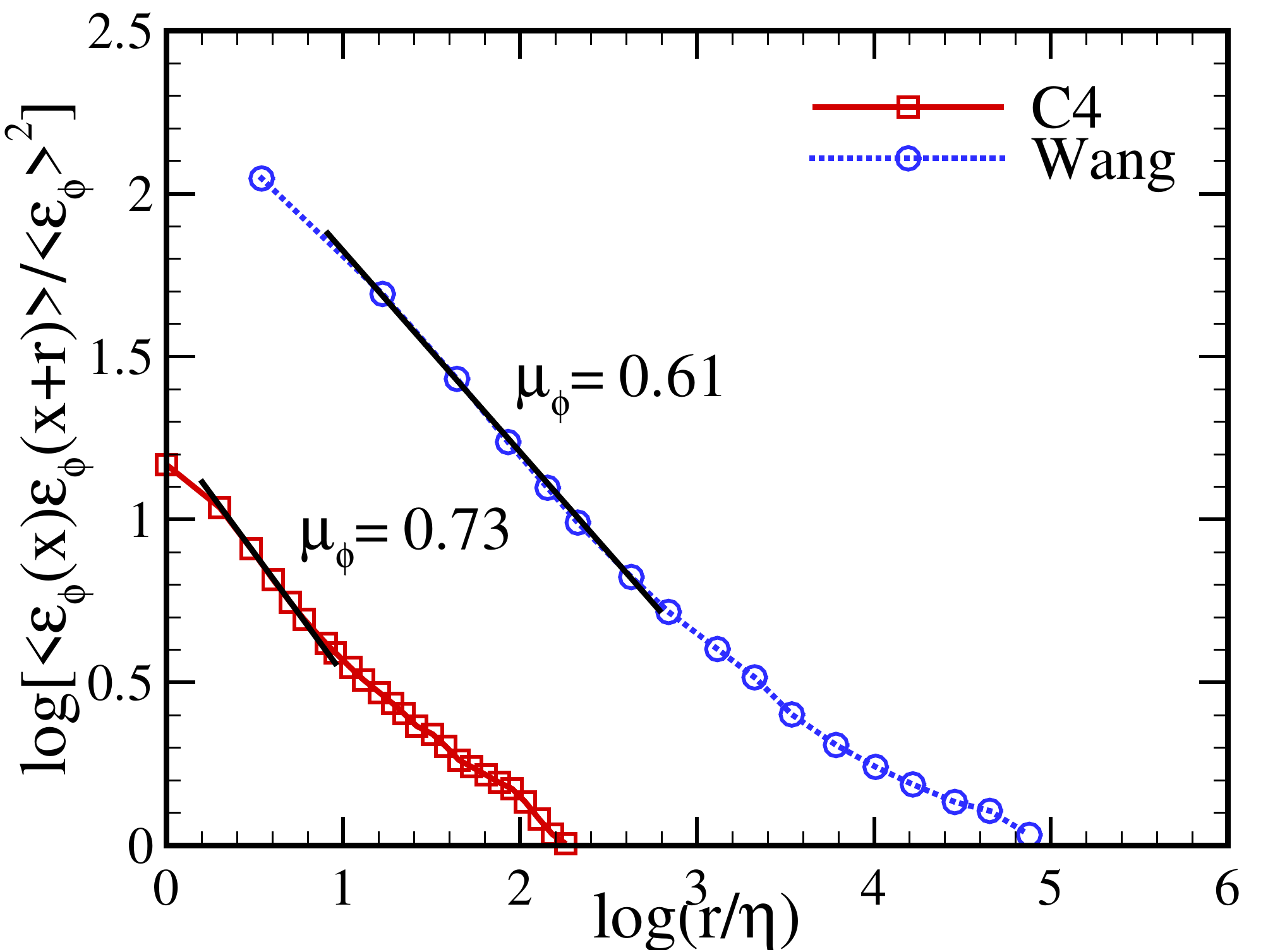}}
\caption{(color online). Auto correlation of scalar dissipation rate, as a function of $r/\eta$, where the squares and
circles are for the data from C4 and [30], respectively.}
\label{fig:fig19}
\end{figure}

Another discriminative issue involves the scalar intermittency in dissipative range. A common used method for quantifying
this intermittency is to compute the so-called intermittency parameter, $\mu_\phi$, through the auto correlation of scalar
dissipation rate, namely,
\begin{equation}
\langle\epsilon_\phi(\textbf{x})\epsilon_\phi(\textbf{x}+\textbf{r})\rangle \sim r^{-\mu_\phi}.
\end{equation}
Fig.~\ref{fig:fig19} presents a log-log plot of the auto correlations of $\epsilon_\phi$ from C4 and [30],
as functions of the normalized separation distance $r/\eta$. It shows that the values
of $\mu_\phi$ from the compressible and incompressible flows are $0.73$ and $0.61$, respectively. This means
that compared with its incompressible counterpart, the scalar dissipation field in the compressible turbulent mixing
is more intermittent by virtue of the contribution from shock waves.

\section{SUMMARY AND CONCLUSIONS}

In this paper, we systematically studied the effects of the Schmidt number on passive scalar transport in compressible
turbulence. The simulations were solved numerically by adopting a hybrid approach of a seventh-order WENO scheme for shock
regions, and an eighth-order CCFD scheme for smooth regions outside shocks. Large-scale predominant compressive forcing was added
to the velocity field for reaching and maintaining a statistically stationary state. The simulated flows were divided
into two groups. One was used to explore the scalar with low molecular diffusivity in low $Re$ flows, where the Schmidt number
$Sc$ was decreased from $25$ to $1$, and the Taylor mircoscale Reynolds number $Re_\lambda$ was around $35$. The other was
addressed the scalar with high molecular diffusivity in high $Re$ flows, where $Sc$ was decreased from $1$ to $1/25$, and
$Re_\lambda$ was around $210$. Our results showed that in both groups the ratio of the mechanical to scalar
timescales increases as $Sc$ decreases. As an alternative to $r_\phi$, $f_\phi$ was found to fall monotonously when the
product of $Re_\lambda$ and $Sc$ increases.

In the inertial-convective range of $L^{-1}_\phi\ll k\ll \eta^{-1}$, the scalar spectrum seems to obey the $k^{-5/3}$ power law,
especially for the $Sc=1$ flows. Besides, the tendencies of the scalar spectra to grow and fall with wavenumbers between the
inertial-convective
and dissipative ranges are reinforced by the increase and decrease in $Sc$, respectively. For the 1D scalar
spectrum, the spectral bump becomes more visible as $Sc$ increases, and the related OC constant and its 3D counterpart
satisfy the relation $C_{1\phi}=3C_\phi/5$. Further, the scalar spectrum in the viscous-convective range
of $\eta^{-1}\ll k\ll \eta_B^{-1}$ from the $Sc=25$ flow follows the $k^{-1}$ power law, while that in the inertial-diffusive
range of $\eta_{OC}^{-1}\ll k\ll \eta^{-1}$ from the $Sc=1/25$ flow shows a $k^{-17/3}$ scaling. At large scales, the
scaling constant computed by the second-order structure function of scalar increment can be approximately described using an
asymptotic formulation developed for incompressible turbulence. Simultaneously, the one computed by the mixed third-order
structure function of velocity-scalar increment shows that the dependence on the Reynolds number is negligible, and the
effect of compressibility makes it smaller than the $4/3$ value from incompressible turbulence. In addition, this
scaling constant has an increasing tendency when $Sc$ grows.

At small amplitudes, the one-point PDF of scalar fluctuations collapses to the Gaussian distribution, whereas at large
amplitudes it is sub-Gaussian, exhibiting as decay more quickly than Gaussian. For the one-point PDF of
scalar gradient, the convex PDF tails are much longer than Gaussian, indicating strong intermittency. In both
high and low $Re$ flows, the PDF tails on each side become broader as $Sc$ increases. Furthermore, for the $Sc=1$ flows
a notable increase in $Re_\lambda$ leads the PDF tails to be significantly wide, implying that the growth
in the Reynolds and Schmidt numbers will enhance the events of extreme scalar oscillations at small scales. At large
scales, the behavior of the two-point PDF of scalar increment resembles the one-point PDF
of scalar fluctuation, while at small scales, it is similar to that of scalar gradient.
In terms of scalar dissipation, it occurs more at large magnitudes when $Sc$ grows.

The contour shows that in the highest $Sc$ flow the scalar field rolls up and gets sufficiently mixed, whereas in the lowest
$Sc$ flow it loses the small-scale structures because of high molecular diffusivity, and leaves
the large-scale, cloudlike structures. A further study on the contours of scalar advection and dissipation finds
that in the highest $Sc$ flow, streamers with extreme values are small in scale and are distributed approximately randomly in space,
whereas in the lowest $Sc$ flow only the large-scale, cloudlike structures exist. In certain ranges, the spectral
densities of scalar advection and dissipation seem to have the $k^{-5/3}$ power law. This indicates that for
the transport of scalar fluctuations in compressible turbulent mixing, the advection and dissipation other than
the scalar-dilatation coupling may follow the Kolmogorov picture. By computing the compensated spectra
of scalar relative to the kinetic energy spectrum, it is confirmed that the scalings of $k^{-1}$ and $k^{-17/3}$ are
defined for the scalar spectra in the viscous-convective and inertial-diffusive ranges, respectively.
It then shows that at high wavenumbers, the magnitudes of spectral coherency in both the highest and lowest $Sc$ flows
decay faster than $k^{-2/3}$, which is not similar to the prediction from classical theory. Finally, the
comparison with incompressible results displays that the scalar in the $Sc=1$ compressible flow lacks a conspicuous
bump structure in its spectrum near the dissipative range; however, it is more intermittent in this range.

In summary, the above findings reveal that the change in the Schmidt number has pronounced influence
on the small-scale statistics and field structure of passive scalar in compressible turbulence. Besides, although the
turbulent Mach number used in current study is not very high, the effect of compressibility still affects scalar mixing in
certain respects. A deeper investigation on this topic under much higher Reynolds and Schmidt numbers will be
carried out in the near future. Also, we will examine the effects of Mach number and forcing scheme on compressible
turbulent mixing.

\section{ACKNOWLEDGMENTS}

The author thanks Dr. J. Wang for many useful discussions. This work was supported by the National Natural
Science Foundation of China (Grants 91130001 and 11221061), and the China Postdoctoral Science Foundation
Grant 2014M550557. Simulations were done on the TH-1A supercomputer in Tianjin, National Supercomputer Center
of China.

\end{document}